\documentclass[12pt]{iopart}
\pdfoutput=1
\usepackage{graphicx}
\usepackage{dcolumn}

\usepackage{amssymb}
\newcommand\dif{\mathop{}\!\mathrm{d}}
\renewcommand{\vec}[1]{\mathbf{#1}}
\newcommand{\pdagger}{{\phantom{\dagger}}}

\usepackage{siunitx}
\usepackage{multirow,bigdelim,dcolumn,booktabs}
\usepackage{bm}
\usepackage{dsfont}
\usepackage{mwe}
\usepackage{comment}
\usepackage[dvipsnames]{xcolor}  

\usepackage{cite}

\definecolor{darkgreen}{rgb}{0.0, 0.5, 0.0}

\usepackage[english,capitalise]{cleveref}
\Crefname{equation}{equation}{equations}
\Crefname{figure}{figure}{figures}
\Crefname{tabular}{Tab.}{Tabs.}
\usepackage{tikz}
\usepackage[utf8]{inputenc} 
\usepackage[T1]{fontenc} 
\usepackage[arrow, matrix, curve]{xy}
\usetikzlibrary{positioning,angles,calc,intersections,quotes,arrows.meta}  
\begin{document}

\title{Dynamics of spin relaxation in nonequilibrium magnetic nanojunctions}

\author{Rudolf Smorka}
\address{Institute of Physics, Albert-Ludwig University Freiburg, Hermann-Herder-Strasse 3, 79104 Freiburg, Germany}

\author{Michael Thoss}
\address{Institute of Physics, University of Freiburg, Hermann-Herder-Strasse 3, 79104 Freiburg, Germany}

\author{Martin \v{Z}onda}
\address{Department of Condensed Matter Physics, Faculty of Mathematics and Physics, Charles University, Ke Karlovu 5, Praha 2 CZ-121 16, Czech Republic}

\date{\today}


\begin{abstract}

We investigate nonequilibrium phenomena in magnetic nano-junctions using a numerical approach that combines classical spin dynamics with the hierarchical equations of motion technique for quantum dynamics of conduction electrons. Our focus lies on the spin dynamics, where we observe non-monotonic behavior in the spin relaxation rates as a function of the coupling strength between the localized spin and conduction electrons. Notably, we identify a distinct maximum at intermediate coupling strength, which we attribute to a competition that involves the increasing influence of the coupling between the classical spin and electrons, as well as the influence of decreasing local density of states at the Fermi level. Furthermore, we demonstrate that the spin dynamics of a large open system can be accurately simulated by a short chain coupled to semi-infinite metallic leads. In the case of a magnetic junction subjected to an external DC voltage, we observe resonant features in the spin relaxation, reflecting the electronic spectrum of the system. The precession of classical spin gives rise to additional side energies in the electronic spectrum, which in turn leads to a broadened range of enhanced damping in the voltage.

\end{abstract}

\section{Introduction}
The ever-growing interest in magnetic nanostructures~\cite{nanomagnetism2006,nanomagnetism2017,wiesendanger2018atomic,barman2020magnetization}, driven primarily by advancements in nano-device fabrication, has, in recent years, motivated the reinvestigation of various fundamental magnetic phenomena essential for real-world applications as well as basic research. 
These include issues such as spin relaxation~\cite{barman2020magnetization,barman2018spin,sayad2015spin,sayad2016relaxation,ibanez2018spin,tenberg2019electron,mondal2016relativistic,wieser2013comparison,Pletyukhov2010}, inertial dynamics~\cite{Olive2015,sayad2016inertia,bajpai2019time,neeraj2021inertial,bhattacharjee2012atomistic}, spin fluctuations~\cite{yang2014two,glasenapp2014spin,rice2016revealing,roy2015interacting,swar2021detection}, externally driven magnetization dynamics~\cite{ralph2008spin,fransson2008switching,filipovic2013spin,fransson2014electrical,hammar2016time,hammar2017transient,filipovic2016photon,filipovic2018shot,hammar2018dynamical,hammar2019,makhfudz2020nutation,neeraj2021inertial,Carva_book2017,Rahman2021}, 
and previously unrecognized torques~\cite{stahl2017anomalous,bajpai2020spintronics,elbracht2020topological,elbracht2021long}.
These studies are often directly applicable to the investigation of dynamical properties of single-molecule magnets~\cite{Benelli2015,bogani2008molecular,leukona2021}, magnetic impurities embedded in metallic hosts~\cite{delgado2015emergence,Benelli2015,Kaminski2001,scott2010,baran2021subgap,taranko2019transient,Krivenko2019}, or even larger ferromagnetic systems whose dynamics can be represented by a single macrospin~\cite{xiao2005macrospin,sayad2012macrospin,ralph2008spin}. They also provide crucial insights into the dynamics of more complex magnetic structures \cite{Petrovic18,bajpai2019time,elbracht2021long,eschenlohr2020spin,neeraj2021inertial,serrate2010imaging,vedmedenko2020,evers2020advances} and the general interplay between localized magnetic moments and conduction electrons~\cite{brataas2012current,sbiaa2017recent,cai2021survey,evers2020advances} relevant for spintronics applications~\cite{wolf2001spintronics,zutic2004spintronics,hirohata2020review,petrovic2021prx}.    

This interplay has already been addressed using a wide range of methods. 
Significant results have been obtained through fully quantum-mechanical approaches, including exact diagonalization~\cite{tsunetsugu1997ground,Sarkka2011,Prelovek2013}, the time-dependent density-matrix renormalization group~\cite{Cazalilla2002,sayad2016relaxation,stahl2017anomalous,petrovic2021prx}, nonequilibrium Green's functions (NEGF) \cite{haug2008quantum,kalitsov2009spin,Rungger2020,Tang2021} or \emph{ab initio} calculations~\cite{Antropov1995,Antropov1996,Kunes2002,Ebert2011,Nikolic2018First,Carva_book2017}. 
Nevertheless, because of the complexity of the problem, classical methods continue to play a crucial role in this research area.  
Arguably, the most widely used approaches are based on the Landau-Lifshitz-Gilbert (LLG) equation~\cite{Prohl2001,bertotti2009nonlinear}, and have been applied to macroscopic, micromagnetic and even atomic length scales~\cite{tatara2008microscopic,Leliaert2019,bertotti2009nonlinear,skubic2008method,Spirit2019,Evans2018}, encompassing a wide variety of interactions, anisotropies and novel spin structures~\cite{bertotti2009nonlinear,koshibae2018theory,kim2019skyrmions,Leliaert2019}.

Both fully quantum-mechanical and fully classical approaches possess strengths as well as limitations. For example, quantum methods are typically restricted to analyzing small systems or short time scales. On the other hand, the LLG equation offers a wider range of applicability. However, when using the LLG equation to describe realistic setups, it often necessitates the inclusion of phenomenological terms and the need to fit its parameters to experimental results or extract them from \emph{ab initio} calculations ~\cite{oogane2006magnetic,zhang2020ultralow,Kumar2018}.

To partially alleviate some of these limitations and to bridge the quantum and classical methods the LLG equation has been recently combined with steady-state NEGF (NEGF+LLG) \cite{chatterjee2012impact,lee2013r,xie2017materials,Bostrom2019} and its time-dependent extensions (TD-NEGF+LLG) \cite{Petrovic18,bajpai2019time,bajpai2020spintronics,suresh2020magnon}.

The (TD-)NEGF+LLG methods and their equivalents incorporate both quantum and classical degrees of freedom and, as such, belong to a broader class of hybrid quantum-classical (QC) approaches~\cite{Stock1997,elze2012linear,stock2005classical,bellonzi2016assessment,kapral1999mixed,crespo-otero2018recent,Zonda2019,zonda2019gapless,smorka2020electronic}. 
QC approaches which do not introduce any additional damping or torque terms and do not rely on further approximations~\cite{elze2012linear,sayad2015spin,sayad2016inertia,elbracht2020topological,elbracht2021long} are suitable for the investigation of principal problems related to relaxation processes due to the electron-spin interactions ~\cite{sayad2015spin,sayad2016inertia,elbracht2020topological,elbracht2021long}.

In this work, we utilize a variation of these techniques, namely a quantum-classical equations of motion (QC-EOM) approach for open quantum systems~\cite{bajpai2019time,smorka2022non}. 
QC-EOM is a QC method which combines the equations of motion for the classical spins with the hierarchical equations of motion technique for the conduction electrons~\cite{tanimura2020numeric,batge2021nonequilibrium}. Its advantage is that, in the case of noninteracting fermions and when focusing on single-particle properties~\cite{han2018exact}, the hierarchy terminates exactly at the second tier, respectively, first tier in the case of wide-band limit~\cite{zheng2007time,jin2008exact,croy2009propagation,zhang2013first,popescu2016efficient,leitherer2017simulation}.~\footnote{Note, that the expectation values of many-particle operators might require a higher truncation tier as discussed in detail Refs.~\cite{han2018exact,schinabeck2020hierarchical}} 
The method is therefore numerically exact even far away from equilibrium, allows one to reach long simulation times, and does not contain additional approximations or phenomenological terms that could distort analysis of the relaxation processes. 

We use QC-EOM to study the spin dynamics, in particular relaxation processes, of a single classical spin embedded in a chain of conduction electrons controlled by an external voltage bias. We show that the effective damping and, therefore, also the relaxation rates of the classical spin are nonmonotonic functions of spin-electron coupling and that they are strongly affected by external voltage with clear resonant features reflecting the static as well as dynamic electronic spectrum of the system. We analyze and discuss some important differences between our results and the results obtained for an equivalent setup in the previous work~\cite{sayad2015spin}. 
Furthermore, we demonstrate that the transient dynamics of a long isolated central system can be modeled by a short open one, i.e., a short chain of several sites coupled to reservoirs. We also discuss the usability of the simpler LLG approach in different regimes of the system.  

This paper is organized as follows.
In section~\ref{sec:Model} we define the model of a quantum-classical magnetic nano-junction.
We describe the QC-EOM formalism for open quantum-classical systems in section~\ref{sec:Method}.
The results for the spin dynamics are presented in section~\ref{sec:Results}.
First, we investigate in section~\ref{sec:Closed} the spin relaxation dynamics in an isolated system, described by a single-impurity Kondo chain with a classical spin.
We then in section~\ref{Sec:ShortChain} show that the dynamics of a long isolated system can be simulated by a short system coupled to semi-infinite leads.
In section~\ref{sec:DCDependence} we introduce a finite voltage drop to the system and investigate the current-driven spin dynamics of the classical spins. We then further elaborate on some of the findings in section~\ref{Sec:MJ} where we focus on a system weakly coupled to the leads which models a simplified magnetic nanojunction. Section~\ref{sec:Summary} summarizes our findings. Some more technical aspects are outlined in the appendices.   

\section{Model of Hybrid Magnetic Nano-Junction}\label{sec:Model}

\begin{figure}[!h]
	\centering	\includegraphics[width=0.7\textwidth]{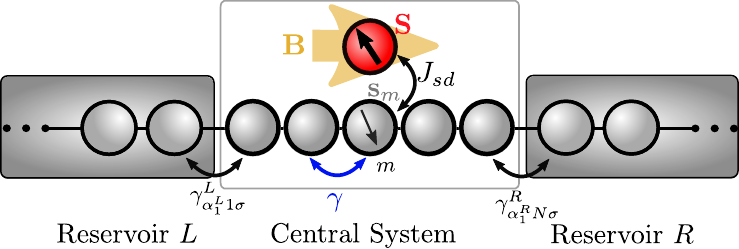}
	\caption{Schematic setup of a quantum-classical magnetic nano-junction. The central system is modeled as a one-dimensional electronic tight-binding chain with a classical spin, and it is coupled to reservoirs $L$ and $R$ of non-interacting electrons.}\label{fig:SchemticOpen}
\end{figure}
We consider a magnetic nano-junction consisting of a central hybrid quantum-classical system~\cite{elze2012linear,sayad2015spin} coupled to reservoirs of non-interacting electrons. The central part is modeled by a one-dimensional electronic tight-binding chain, in which electrons interact locally with a classical spin positioned at the center of the chain (see figure~\ref{fig:SchemticOpen}). The Hamiltonian of the quantum sector comprises three contributions and reads 
\begin{equation}
	H_\mathrm{O}(t) = H_\mathrm{C}(t) + \sum_{\ell\in\{L,R\}}\left(H_\ell + H_{\mathrm{C}\ell}\right),
    \label{eq:4_Hamiltonian_Quantum_Open}
\end{equation}
where the central electronic chain is described by $H_\mathrm{C}$, the fermionic reservoirs $\ell=\{L,R\}$ by $H_\ell$, and the coupling between the chain and reservoir $\ell$ by $H_{\mathrm{C}\ell}$. 
The tight-binding chain coupled to a classical spin is modeled by the (classical) single-impurity Kondo model (with $s$-$d$-type interaction)
\begin{eqnarray}
	H_\mathrm{C}(t) &= &-\gamma \sum_{\langle j,j'\rangle,\sigma} \left( c_{j\sigma}^\dagger c_{j'\sigma}^\pdagger + \mathrm{h.c.}\right) + \overline{\mu}\sum_{j,\sigma}c_{j\sigma}^\dagger c_{j\sigma}^\pdagger\nonumber\\ &&+\frac{J_{sd}}{2}  \sum_{\sigma\sigma'}c^\dagger_{m\sigma}(\vec{S}_m(t)\cdot \bm{\sigma})_{\sigma\sigma'}^\pdagger c^\pdagger_{m\sigma'}.
\label{eq:Hamiltonian_Quantum_Closed}
\end{eqnarray}
The first term in equation~(\ref{eq:Hamiltonian_Quantum_Closed}) is the kinetic energy associated with electron hopping between neighboring sites. 
Here, $c_{j\sigma}^\dagger$ and $c_{j\sigma}^\pdagger$ denote the fermionic creation and annihilation operators with spin $\sigma=\{\uparrow,\downarrow\}$ at site $j$ and $\gamma$ is the hopping integral.
In the analysis below, we fix $\gamma = 1$ meaning that we are using $\gamma$ as the energy unit and time is in units of $\gamma^{-1}$. All relevant physical constants are absorbed into the model parameters.

The second term controls the electronic filling via electrochemical potential $\overline{\mu}$. Here, we assume that the system is small enough that its equilibrium electrochemical potential can be set externally, e.g., by an auxiliary gate electrode, which does not contribute to the charge and spin transport. If not stated otherwise,  $\overline{\mu}$ is taken to zero which sets the half-filling condition.

The last term describes the local coupling of a localized spin $\vec{S}_m(t)\in \mathds{R}^3$ to the spin polarization of the electrons at the chain site $m$, where $\bm{\sigma}$ denotes the vector of Pauli matrices. Note that $H_\mathrm{C}(t)$ acquires a time dependence through $\vec{S}_m(t)$, as discussed in detail below.  

The electronic reservoirs and their coupling to the central chain is modeled by
\begin{eqnarray}
	H_\ell &=& \sum_{\alpha\sigma} \varepsilon^\ell_{\alpha\sigma}d_{\ell\alpha\sigma}^\dagger d_{\ell\alpha\sigma}^\pdagger,\\
	H_{\mathrm{C}\ell} &=& \sum_{\alpha j\sigma} \left(\gamma^\ell_{\alpha j \sigma}c_{j\sigma}^\dagger d_{\ell\alpha\sigma}^\pdagger +\mathrm{h.c.}\right),
\end{eqnarray}
respectively.
Here, $d_{\ell\alpha\sigma}^\dagger$ and $d_{\ell\alpha\sigma}^\pdagger$ denote the fermionic creation and annihilation operators of the reservoirs.
Further, $\varepsilon_{\alpha \sigma}^\ell$ is the single-particle energy of state $\alpha$ with spin $\sigma$ in reservoir $\ell$, and $\gamma^\ell_{\alpha j\sigma}$ is the coupling constant, connecting site $j$ at the edge of the central chain to state $\alpha$ in the reservoir $\ell$ (see figure~\ref{fig:SchemticOpen}). 
This constraint implies that the system-reservoir coupling matrix is nonzero only at the interface sites. For a linear chain, these are $j=1$ for $\ell=L$ and $j=N$ for $\ell=R$. 
Here, the spin-flip processes between the reservoirs and the central system are forbidden $\gamma_{\alpha j\sigma\sigma'}^\ell = 0$ (where $\sigma \neq \sigma'$). 
Therefore, the coupling is diagonal in the basis of $\sigma_z$. 
The reservoirs are assumed to be in chemical and thermal equilibrium, with chemical potential $\mu_\ell$ and inverse temperature $\beta_\ell$.
 
\section{Quantum-Classical Equations of Motion}\label{sec:Method}

We employ the QC-EOM technique, which consists of a set of quantum equations of motion for the conduction electrons coupled with classical equations of motion for the classical spin. The time-evolution of the quantum sector of the magnetic junction is governed by the Liouville-von Neumann equation for the electronic density matrix
\begin{equation}
	i\frac{\partial }{\partial t} \rho(t) = [H(t),\rho(t)].\label{eq:EOM_Quantum_Closed}
\end{equation}
When considering an isolated system, that is in the absence of fermionic reservoirs, $H(t)$ in equation~(\ref{eq:EOM_Quantum_Closed}) is identical to $H_\mathrm{C}(t)$ from equation~(\ref{eq:Hamiltonian_Quantum_Closed}).
 
In the presence of fermionic reservoirs, $H(t)$ is identified with $H_\mathrm{O}(t)$ from equation~(\ref{eq:4_Hamiltonian_Quantum_Open}). 
A system of equations of motion for the reduced density matrix $\rho_\mathrm{C}(t) = \mathrm{tr}_{L+R} \{\rho(t)\}$ is obtained by tracing out the reservoir degrees of freedom from $\rho(t)$ (where $\mathrm{tr}_{X}\{\cdot\}$ denotes the partial trace over the $X$ sub-system).
Therefore, the dynamics of the magnetic junction is reduced to the central part only.  
To describe the dynamics of the magnetic junction, we use a hierarchical equations of motion approach \cite{tanimura2020numeric,batge2021nonequilibrium,smorka2022non}.
For the case of non-interacting fermions and single-particle properties studied in this paper, the hierarchy of equations of motion for the auxiliary density matrices terminates at the second tier exactly~\cite{zheng2007time,jin2008exact,zhang2013first,croy2009propagation,popescu2016efficient,han2018exact,leitherer2017simulation,schinabeck2020hierarchical}. 
The equation of motion for the zeroth-tier auxiliary density matrix  $\rho_\mathrm{C}$, that is the reduced density matrix of the central system
\begin{equation}
	i\frac{\partial }{\partial t} \rho_\mathrm{C}(t) = [H_\mathrm{C}(t),\rho_\mathrm{C}(t)] + i \sum_\ell \left(\Pi^\dagger_\ell(t) + \Pi_\ell^\pdagger(t) \right),\label{eq:RhoOpenEOM}
\end{equation}
contains a unitary time-evolution under the (time-dependent) Hamiltonian $H_\mathrm{C}(t)$ of the central system. The second term on the right hand side of equation~(\ref{eq:RhoOpenEOM}) generates dissipation, a non-unitary time-evolution due to the coupling of the central system to the fermionic reservoirs.

The dissipation operator is defined in terms of first-tier
auxiliary density matrices $\Pi_\ell$ (called \emph{current matrices} for brevity), which can be expressed via time-dependent nonequilibrium Green's functions 
\begin{equation}
	\Pi_\ell (t) = \int_{-\infty}^t \dif \tau [G^>(t,\tau)\Sigma_\ell^<(\tau,t)- G^<(t,\tau)\Sigma_\ell^>(\tau,t)],
\label{eq:CM}
\end{equation}
with lesser and greater Green's functions $G^\lessgtr$ and $\Sigma^\lessgtr_\ell$ being the lesser and greater tunneling self-energies due to presence of reservoir $\ell$ \cite{rahman2018non,leitherer2017simulation,popescu2016efficient,haug2008quantum}. For further details, see~\ref{App:CurMat}.
We use the wide-band approximation, where the line-width functions are energy independent and read
\begin{equation}
	\Gamma^\ell_{jk}(\varepsilon) \equiv 2\pi\sum_{\alpha\sigma} \gamma^\ell_{\alpha j\sigma} (\gamma^\ell_{\alpha k\sigma'})^*\delta(\varepsilon-\varepsilon_{\alpha\sigma}^\ell)=\Gamma^\ell_{jk}\delta_{\sigma\sigma'}\delta_{jk}.
	\label{eq:Gamma}
\end{equation} 
Given the one-dimensional geometry employed throughout this work, the coupling is finite at the interface sites $j=1,N$ only.
The previously stated constraints on $\gamma_{\alpha j\sigma}^\ell$ due to the considered geometry and spin conservation at the interface lead to an analogous matrix-structure of $\Gamma_{jk}^\ell(\varepsilon)$ as that of $\gamma_{\alpha j\sigma}^\ell$.\\

By utilizing the Padé-decomposition of the reservoir Fermi-Dirac distribution \cite{hu2010communication} 
\begin{equation}
	f(\varepsilon) \approx \frac{1}{2}-\frac{1}{\beta}\sum_{p=1}^{N_p}\eta_p \Bigg(\frac{1}{\varepsilon-\chi_{p\ell}^-} +\frac{1}{\varepsilon-\chi^+_{p\ell}}\Bigg),\label{eq:PadeDec}
\end{equation} 
where $\chi_{p\ell}^\pm (t) = \mu_\ell(t)\pm i\xi_p \beta^{-1}$ and $\eta_p$ are Padé coefficients of the $N_p$ poles of Padé-expansion,
the current matrices take the following form 
\begin{equation}
	\Pi_\ell(t) = \frac{1}{4}(\mathds{1}-2\rho_\mathrm{C}(t))\Gamma_\ell+\sum_{p=1}^{N_p} \Pi_{\ell,p}(t),\label{eq:AuxiliaryRhoDef}
\end{equation}
with Pad\'{e}-resolved auxiliary current matrices $\Pi_{\ell,p}$ following the equation of motion~\cite{croy2009propagation,popescu2016efficient,leitherer2017simulation}
\begin{equation}
	i\frac{\partial }{\partial t}\Pi_{\ell,p}(t) = \frac{\eta_p}{\beta}\Gamma_\ell + 
	 \left(H_\mathrm{C}(t)-\frac{i}{2}\Gamma-\chi_{\ell, p}^+(t)\mathds{1}\right)\Pi_{\ell,p}(t).\label{eq:PadeAuxiliaryRhoEOM}
\end{equation}
Here, $\Gamma=\sum_\ell\Gamma_\ell$ denotes the total line-width function. 
From the current matrices $\Pi_\ell$, the charge $I_\ell$ and spin currents $Q^\alpha_\ell$ flowing through the interface between the reservoir $\ell$ and the system can be obtained
\begin{eqnarray}
I_\ell(t) &=& \mathrm{Re}\, \mathrm{tr}\{\Pi_\ell(t)\},\label{eq:DefCurrent}\\
Q^\alpha_\ell(t) &=& \mathrm{tr}\{\sigma_\alpha \Pi_\ell(t)\}.\label{eq:DefSpinCurrent}
\end{eqnarray}
Note, that neither the Padé expansion~\cite{croy2009propagation} nor quantum-classical methods in general are in anyway limited to the here used wide-band limit~\cite{bajpai2019time,bajpai2020spintronics,Petrovic18,suresh2020magnon}. We use it for convenience, in particular to reduce the parameter space and the number of tiers in the hierarchy. In particular, the hierarchy for single-particle operators terminates at the first tier exactly within the wide-band limit \cite{zhang2013first,erpenbeck2018current}. The wide-band limit also allows for a more straightforward analysis of the results. 

The classical sector contains a single localized spin $\vec{S}_m(t)$, and its dynamics is generated by the classical Hamiltonian
\begin{equation}
	\mathcal{H} = J_{sd} \vec{s}_m(t)\cdot\vec{S}_m(t) - \vec{B}(t)\cdot \vec{S}_m(t).\label{eq:Hamiltonian_Classical}
\end{equation}
The classical spin couples to the expectation value of the local conduction electron spin polarization
\begin{equation}
	\vec{s}_m(t) = \frac{1}{2}\mathrm{tr}\{\rho_{m}(t)\bm{\sigma}\},\label{eq:SpinPolarization}
\end{equation}
defined in terms of the reduced density matrix $\rho_m = \mathrm{tr}_{\Lambda\diagdown m}\{\rho \}$ at site $m$, obtained by tracing out all chain degrees of freedom $\Lambda$ except site $m$. 
Furthermore, we assume an external magnetic field $\vec{B}(t)$ acting on the classical spin only, which gives rise to a Zeeman contribution in equation~(\ref{eq:Hamiltonian_Classical}).

Using the extension of classical Poisson-brackets to spin systems~\cite{yang1980generalizations}, the classical spin equation of motion is

\begin{eqnarray}
	\frac{\dif }{\dif t} \vec{S}_m(t) &=& \{\vec{S}_m(t),\mathcal{H}(t)\} =  \vec{S}_m(t)\times\vec{B}^\mathrm{eff}(t),\label{eq:EOM_Classical}\\
	\vec{B}^\mathrm{eff}(t)&=&-\nabla_{\vec{S}_m}\mathcal{H}(t) = -J_{sd}\vec{s}_m(t) + \vec{B}(t).\label{eq:EffectiveField}
\end{eqnarray}
Herein, the local effective field $\vec{B}^\mathrm{eff}$ is obtained from the gradient $\nabla_{\vec{S}_m}$ of the classical Hamiltonian $\mathcal{H}$ with respect to the classical spin. In the following, we assume $\vert\vec{S}_m(t)\vert = S=\mathrm{const.}$ with $S=1$.

The QC-EOM method thus consists of solving the coupled set of equations of motion \cref{eq:RhoOpenEOM,eq:AuxiliaryRhoDef,eq:PadeAuxiliaryRhoEOM} in the quantum sector and simultaneously the classical equation of motion~(\ref{eq:EOM_Classical}), coupled by the $s$-$d$ term in equation~(\ref{eq:Hamiltonian_Quantum_Closed}) and the spin polarization expectation value~(\ref{eq:SpinPolarization}).
We note that the quantum-classical approach employed here is an Ehrenfest-type method \cite{stock2005classical,elze2012linear,bellonzi2016assessment}. The Ehrenfest approach has been
used to study nuclear dynamics in quantum transport in
Refs.~\cite{verdozzi2006classical,metelmann2011adiabaticity}, and, in particular, has been combined
with the hierarchical equations of motion approach in
Ref.~\cite{erpenbeck2018current}.

\section{Results}\label{sec:Results}

Employing QC-EOM, we investigate the spin relaxation dynamics. We start our analysis with long isolated central system and then compare the dynamics to results obtained for the open system, i.e., short chain coupled to reservoirs. After demonstrating that short-time dynamics of long isolated systems and an open system are equivalent, we investigate the influence of external driving on the dynamics of the central spin. By addressing a junction with single and five chain points, we discuss the role of the electron states and their polarization on the relaxation.

\subsection{Isolated Central System}\label{sec:Closed}
We first analyze an isolated tight-binding chain with a single classical spin adsorbed at its center. 
An analogous system was addressed by Sayad and Potthoff~\cite{sayad2015spin} who investigated the relaxation dynamics by focusing on the switching time of the classical spin after reversing the external magnetic field. 
In contrast, our emphasis lies in extracting the effective (time-independent) Gilbert damping $\alpha$. The switching and relaxation time is proportional to $\alpha/(1+\alpha^2)$~\cite{kikuchi1956minimum,sayad2015spin}, allowing us to reconstruct its profile from the fitted damping coefficient. Fitting the latter has two main advantages. 
First, it can be extracted at much shorter simulation times, often way before the classical spin can be considered numerically relaxed. Consequently, we can analyze shorter chains in our study. 
Second, with the introduction of a suitable fitting model, determining the effective Gilbert damping does not require an arbitrary criterion for identifying the precise moment when the spin is considered fully relaxed in a numerical simulation. 
This aspect proves particularly significant in regimes characterized by long-lived nutations.  

In order to analyze the dependence of spin relaxation on spin-electron coupling $J_{sd}$, we apply a two-stage switching protocol. In the first stage, the localized spin is set to $\vec{S}_{m0} = \vec{e}_x$ (e.g., prepared by a strong external field pointing to the $x$-direction) and electrons are in the ground state with density matrix  
\begin{equation}
\rho_{0,jk} = \sum_{\alpha=1}^{2L} U_{j\alpha}^\pdagger U_{\alpha k}^\dagger \Theta(\mu-\varepsilon_\alpha).
\end{equation}
Here, $U$ describes the unitary transformation to the eigenstates of the initial quantum Hamiltonian [equation~(\ref{eq:Hamiltonian_Quantum_Closed}), at $t=0$] and $j,k$ are indexing system sites as well as spin-projections. 
We consider the electronic system at half-filling set by electrochemical potential $\overline{\mu}=0$.
The second stage is initiated at time $t=t_0=0$ by a sudden switch of the external magnetic field $\vec{B}\rightarrow \vec{B}'=\mathcal{B}\vec{e}_z$, which drives the classical spin out of its initial orientation towards a new steady state."
\begin{figure}[!ht]
	\centering
	\includegraphics[width=1.0\textwidth]{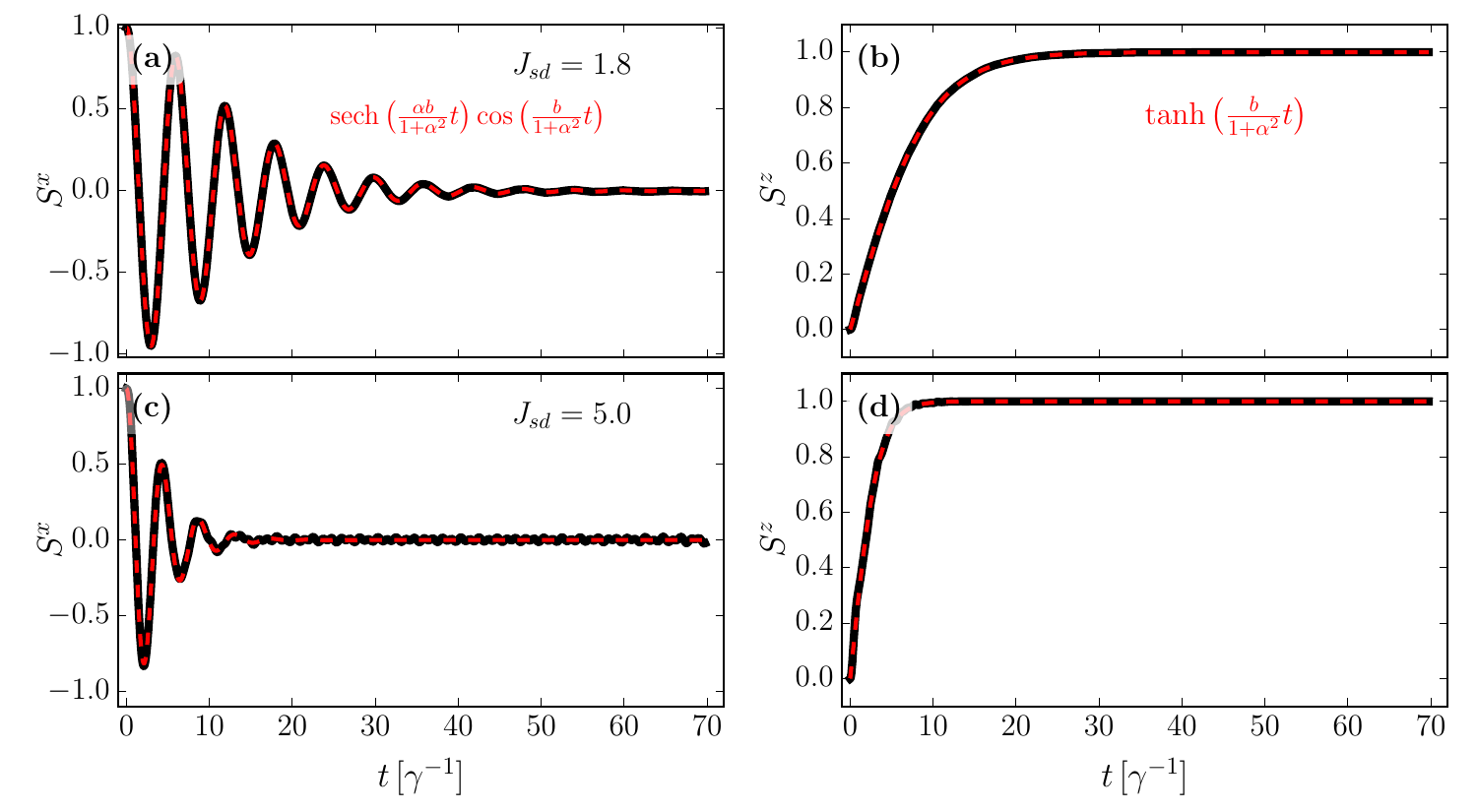}
	\caption{(a),(c) Dynamics of the $S^x$, and (b),(d) $S^z$ components of the classical spin (black lines) for $J_{sd}=1.8$ (a),(b) and $J_{sd}=5.0$ (c),(d) both with $N=151$, and their respective fits to the LLG solution equations~(\ref{eq:FitX}),(\ref{eq:FitZ}) (red dashed lines).}\label{fig:3_DampingJsd_Closed_1}
	\end{figure} 
 \begin{figure}[!ht]
 	\centering
 	\includegraphics[width=0.9\textwidth]{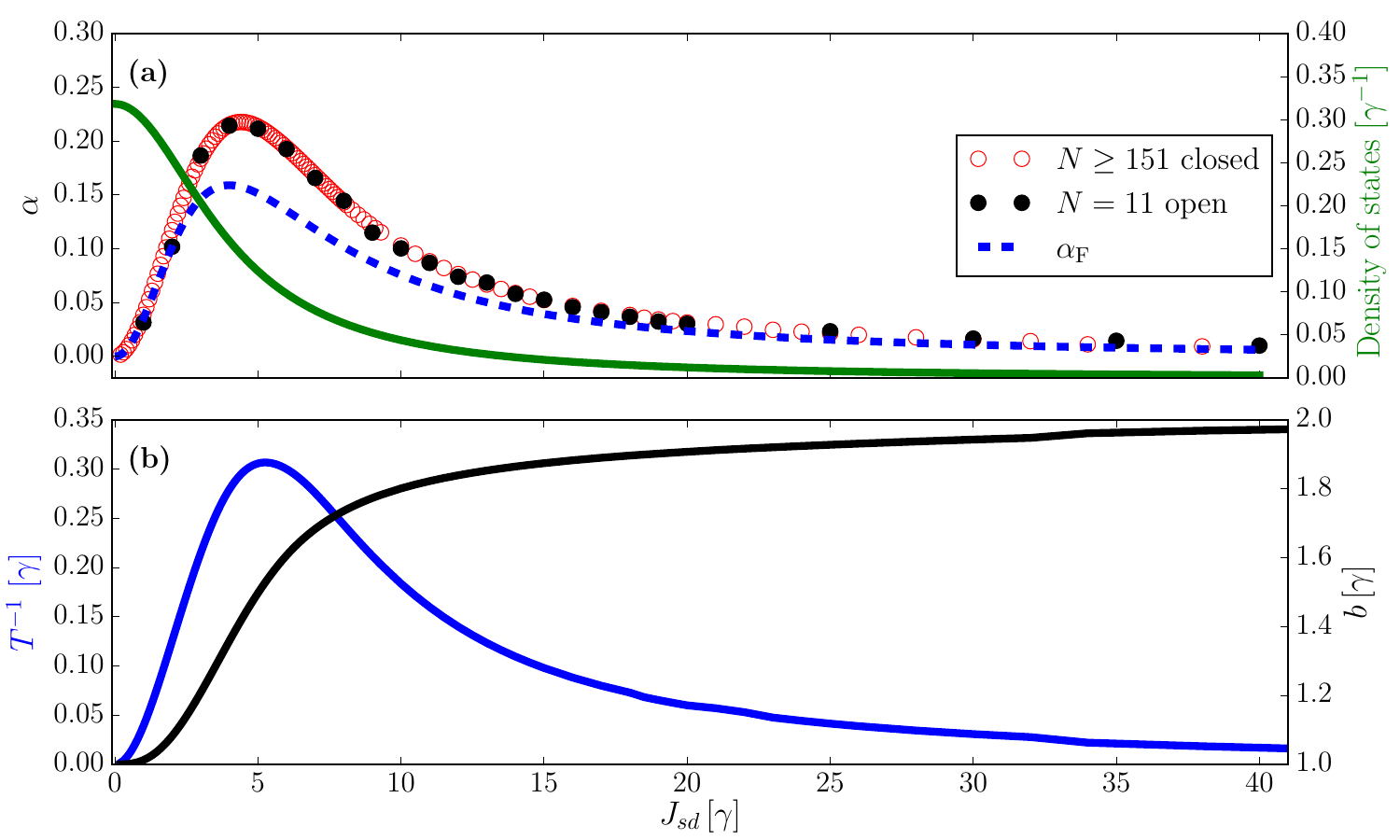}
 	\caption{(a) Left axis: Comparison of the effective Gilbert damping $\alpha$ fitted to the dynamics of long closed system (red empty circles) and the short open system using $\Gamma_{\ell} = 2$(black circles) with approximation $\alpha_F$ from equation~(\ref{eq:gd}) (dashed blue line). Right axis: Local (on-dot) density of states at the Fermi level from equation~(\ref{eq:Depsf}) (green solid line). (b) Left axis: Relaxation rate $T^{-1}$ (blue line) as functions of $J_{sd}$. Right axis: The fitted effective field $b$ (black line).}
 	\label{fig:3_DampingJsd_Closed_2}
 \end{figure}
In the following analysis we assume $B=\mathcal{B}/\gamma=1$ and use odd-numbered chains of length $N\geq151$ with the classical spin coupled to the central position $m=(N-1)/2$. Figure~\ref{fig:3_DampingJsd_Closed_1} shows the dynamics of $S^x$ and $S^z$ for $J_{sd}=1.8$ (a,b) and $J_{sd}=5$ (c,d) plotted with solid black lines. 

For further analysis, we also consider the approximation of these results using the LLG equation~\cite{evans2014atomistic} 
for the single spin in a constant magnetic field $\vec{b}$ pointing to the $z$ direction (its components are $(0,0,b)$)  
\begin{equation}
\frac{\dif }{\dif t} \vec{S}(t) = \vec{S}\times \vec{b} + \alpha \vec{S}\times \frac{\dif }{\dif t} \vec{S}(t),
\label{eq:LLG} 
\end{equation}
which has, for our initial conditions, the exact solution:
\begin{eqnarray}
	S^x(t)&=&\text{sech}\left(\frac{\alpha b}{1+\alpha^2} t\right)\cos\left(\frac{b}{1+\alpha^2} t\right),\label{eq:FitX}\\
	S^z(t)&=&\tanh\left(\frac{\alpha b}{1+\alpha^2}\,t\right),\label{eq:FitZ}
\end{eqnarray}
where both the effective (time independent) Gilbert damping $\alpha$ and the effective magnetic field $b$ are fitting parameters. Note that $b$ cannot be identified with the real magnetic field $B_z$. The reason is that the dominant effective precession frequency $\omega_p=b/(1+\alpha^2)$ of the full system depends on $J_{sd}$ due to the geometrical torque as was already discussed in Ref.~\cite{elbracht2020topological}.

We have performed a nonlinear least-squares analysis of $S^x$ (and independently $S^y$) for various $J_{sd}$. Representative fits (red lines) for $J_{sd}=1.8$ and $J_{sd}=5$ are shown in figure~\ref{fig:3_DampingJsd_Closed_1}.
Figure~\ref{fig:3_DampingJsd_Closed_2} shows the fitted effective Gilbert damping $\alpha$  as red empty circles in panel (a) and the fitted effective magnetic field $b$ (black line) as well as the relaxation rate $T^{-1}=\alpha b/(1+\alpha^2)$ (blue line) in panel (b).

From figure~\ref{fig:3_DampingJsd_Closed_2}(a) it is clear that the spin-electron coupling significantly influences the damping and, therefore, also the relaxation times. The overall shape of the $\alpha$-$J_{sd}$ dependence, which contains a broad maximum at $J_{sd}\approx4$, can be qualitatively understood following the approximate formula for Gilbert damping~\footnote{The factor $1/8$ comes from the fact that we define the Hamiltonian with $J_{sd}/2$ and use total spin-unresolved ${\mathcal D}_{\varepsilon_F}$ }
\begin{equation}
	\alpha_F=\frac{\pi}{8}J^2_{sd}{\mathcal D}_{\varepsilon_F}^2,
	\label{eq:gd}
\end{equation}   
which was derived several times in the literature by different methods, see, e.g., Refs.~\cite{sakuma2006microscopic,nunez2008effective,hosho2006voltage,sayad2015spin,sakuma2013gilbert,turek2015nonlocal}. Here ${\mathcal D}_{\varepsilon_F}$ is the local density of the states at the Fermi level ($\varepsilon_F=0$) evaluated at site $m$. This quantity can be obtained by taking advantage of the Green's function formalism~\cite{haug2008quantum,economou2006green}. 
Assuming a fixed classical spin, we can calculate ${\mathcal D}_{\varepsilon_F}$ as the density of states of a non-interacting quantum dot in a magnetic field $J_{sd}\vec{S}/2$ coupled to two semi-infinite chains of non-interacting electrons. 
Here, the hopping between the dot and the chain is $\gamma=1$ and chains can be fully represented by their edge (surface) density of states per spin~\cite{economou2006green,cizek2004theory}
\begin{equation}
	\mathrm{DOS_e}(\varepsilon)=\frac{1}{2\pi\gamma^2}\sqrt{4\gamma^2-\varepsilon^2}.
	\label{eq:DOS}
\end{equation}          
By using the spin-resolved retarded $G^{r}(\varepsilon)$ and advanced $G^{a}(\varepsilon)$ Green functions of the coupled dot, we can express the on-dot density of states at the Fermi level as
\begin{equation}
	{\mathcal D}_{\varepsilon_F}=i\mathrm{Tr}\,\left[G^{r}(\varepsilon_F)-G^{a}(\varepsilon_F)\right]/2\pi.
	\label{eq:Depsf}
\end{equation}	
For further details see Refs.~\cite{cizek2004theory,Zonda2019,zonda2019gapless,smorka2020electronic}. As it is shown in  figure~\ref{fig:3_DampingJsd_Closed_2}(a),  ${\mathcal D}_{\varepsilon_F}$ is a decreasing monotonic function of $J_{sd}$ (solid green line). Its fast decrease reflects the fact that the classical spin acts as an external magnetic field, which splits the, otherwise degenerated, energy level of the dot symmetrically around $\varepsilon=0$ by $\pm J_{sd}/2$. The finite ${\mathcal D}_{\varepsilon_F}$ results from the overlap of the these splitted and broadened levels. The broadening results from the coupling to the rest of the chain acting as two semi-infinite leads. Because the overlap quickly decays with increasing $J_{sd}$ so does the density of the states at the Fermi level. For example, if we would assume a simple Lorentzian broadening functions (i.e., constant $\mathrm{DOS_e}$) the ${\mathcal D}_{\varepsilon_F}$ and, therefore, also $\alpha_F$ would for large $J_{sd}$ decrease as $\sim J^{-2}_{sd}$.         

Consequently, the resulting $\alpha_F$ dependence, plotted by dashed blue line in figure~\ref{fig:3_DampingJsd_Closed_2}(a), contains a broad maximum for $J_{sd}\approx4$, because the increase of $J_{sd}$ cannot compensate the decrease of ${\mathcal D}_{\varepsilon_F}$ above this value. In addition, because the relaxation rate is linear in $b$ which rises from $b/B=1$ at $J_{sd}\ll 1 $ to $b/B\approx 2$ for $J_{sd}\gg 1$, the maximum of $T^{-1}$ is shifted to higher $J_{sd}\approx 5$ when compared to the position of the maximum of effective Gilbert damping ($J_{sd}\approx4$).  

Similar behavior was also observed for the switching time in Ref.~\cite{sayad2015spin}. However, there are some differences. First, the authors of the cited work stated that their switching time does not scale as $1/J^2_{sd}$, as expected for weak $J_{sd}$. This was attributed to finite-size effects (backscattering) and related numerical instabilities due to very long spin-reversal times for small $J_{sd}$ presented in their work. 
In contrast, because relaxation, respectively, switching times, are proportional to $(1+\alpha^2)/\alpha$ we can conclude that in our case the relaxation times scale with $1/J^2_{sd}$ for $J_{sd}\ll1$ where ${\mathcal D}_{\varepsilon_F}$ is approximately constant. 
This confirms the stability of our method in the weak $J_{sd}$ regime.
Second, in Ref.~\cite{sayad2015spin} the extremal point of the switching time was placed at $J_{sd}\approx30$ which is much higher than our result $J_{sd}\approx5$. This difference cannot be accounted for by the difference in the used magnitude of the classical spin ($|S|=1/2$ in Ref.~\cite{sayad2015spin} and $|S|=1$ here) or external magnetic field. The latter does not have a significant effect on the position of this maximum, as we discuss in~\ref{App:BVar}. 

A plausible explanation for this shift of the maximum in Ref.~\cite{sayad2015spin} is the influence of high-frequency oscillations imposed on top of the dominant precession, e.g., nutations~\cite{hammar2017transient,thonig2015gilbert,stahl2017anomalous}. These higher-order oscillations emerge in the dynamics for intermediate and strong $J_{sd}\gtrsim 4$~\cite{sayad2016inertia,stahl2017anomalous}, as discussed in~\ref{App:Frequency}, and are long-lived~\cite{sayad2016inertia}. Therefore, depending on the criterion chosen for the fully relaxed (switched) classical spin in a numerical simulation, they can significantly influence the extracted switching time for strong $J_{sd}$. Our fitting model~(\ref{eq:LLG}) does not take into account these higher-order terms. A clear difference between the fit and the full dynamics at long simulation times can be seen already for $J_{sd}=5$ in figure~\ref{fig:3_DampingJsd_Closed_2}(b). The spin dynamics of the fit (red line) is quickly damped, but small steady oscillations (of different frequency than $\omega_p$) can be seen in the full dynamics (black line) for a very long time in figure~\ref{fig:3_DampingJsd_Closed_1}(c).  Therefore, the results in figure~\ref{fig:3_DampingJsd_Closed_2} for $J_{sd}>4$ should be understood as the analysis of the relaxation of the dominant spin dynamics. Nevertheless, this seems to be the correct approach, as it was already shown in Ref.~\cite{sayad2016inertia} that the quantum-classical models highly overestimate the relaxation time of high-order spin processes, such as nutations, when compared with the results of fully quantum methods.        
 
\subsection{Short chain coupled to leads \label{Sec:ShortChain}} 

\begin{figure}[!h]
	\centering
	\includegraphics[width=0.7\textwidth]{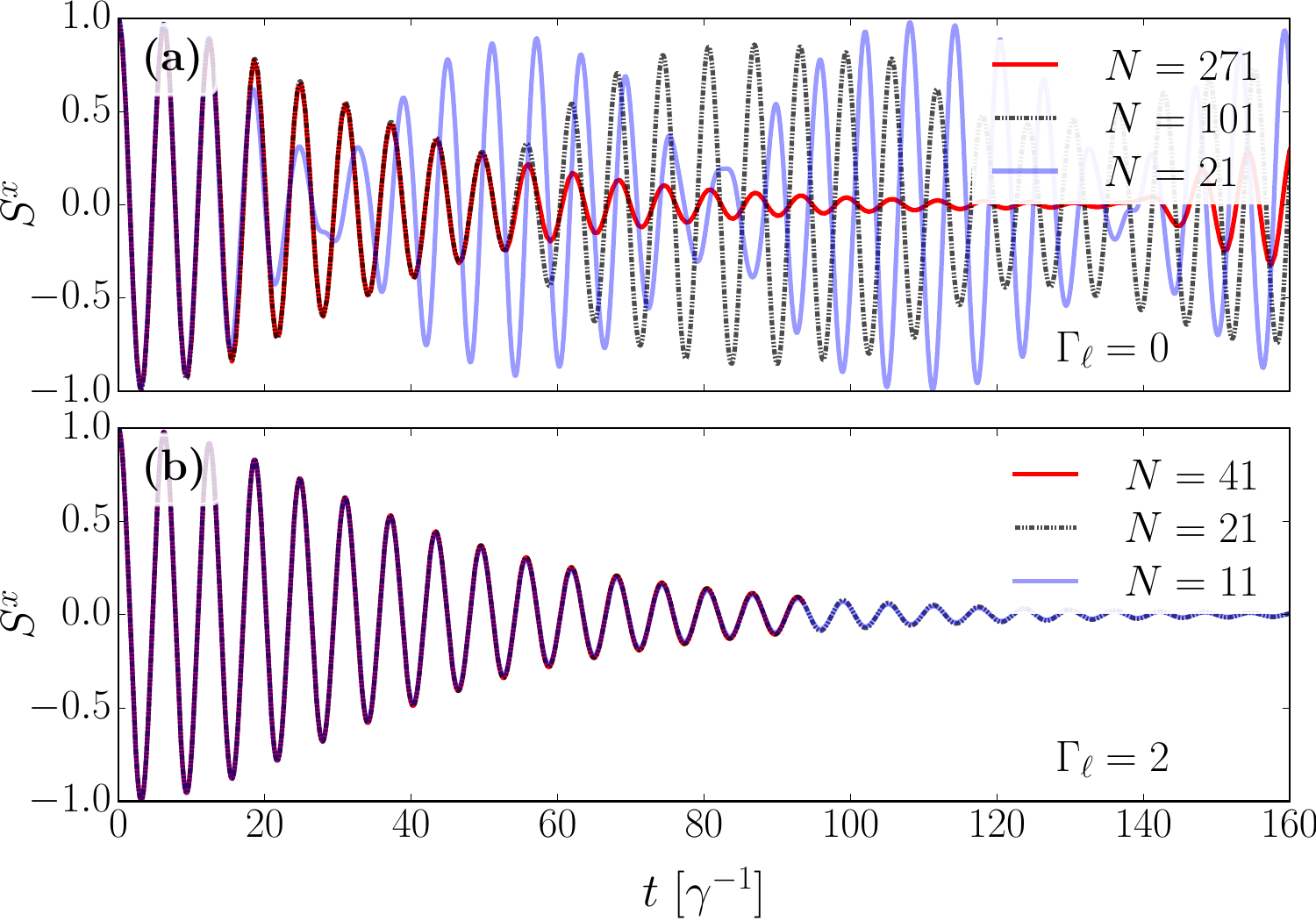}
	\caption{Comparison of spin dynamics of $S^x$ for $\Gamma_\ell = 0$ (a) with systems sizes $N=271$ (solid red line), $N=101$ (black, dotted) and $N=21$ (blue), and dynamics for $\Gamma_\ell = 2$ (b) for $N=41$ (red), $N=21$ (black, dotted) and $N=11$ (blue) at $B=1$, $J_{sd}=1$, $V=0$ and $\beta_\ell = 40$. }\label{fig:4_ComparisonOpenClosed}
\end{figure}

Qualitatively, spin relaxation can be understood as a dissipation of the local nonequilibrium electronic spin excitation into the remaining chain in form of spin-polarized electronic wave-packets~\cite{sayad2015spin,sayad2016relaxation}.
Due to the finite size of the system, these traveling spin wave-packets reflect at the boundaries and after time $\tau=N/(2\gamma)$ interact with the classical spin, leading to recurrences.  
In the above analysis we have always used long enough chains to make the results free of any finite-size effects. 
A counterexample demonstrating recurrences in the spin dynamics for different chain lengths is shown in figure~\ref{fig:4_ComparisonOpenClosed}(a).

Finite-size effects can be mitigated even for short chains by coupling these to a reservoir~\cite{Elbracht2020long,elbracht2021long}. In our case we couple the system to semi-infinite fermionic leads.
Figure~\ref{fig:4_ComparisonOpenClosed} compares spin dynamics in an isolated finite chain of sizes $N=21$ (blue), $N=101$ (black, dotted) and $N=271$ (red) [panel (a)], with that in a magnetic junction ($V=0$) for $\Gamma_\ell = 2$ , $N=11$ (blue), $N=21$ (black, dotted) and $N=41$ (red) [panel (b)]. In both cases, we set $J_{sd}=1$ and $B=1$. The flat band with $\Gamma_\ell=2$ was chosen to approximate the broadening function of the semi-infinite tight-binding chain with $\gamma=1$ around the Fermi level, for which, using the semicircle DOS in equation~(\ref{eq:DOS}), we get $\Gamma_\ell=2\pi\mathrm{DOS}_e(\varepsilon=0)=2$ (i.e., the flat $\mathrm{DOS}_e$ is set to $1/\pi$). Therefore, spin recurrences are strongly suppressed as only a small fraction of electrons is reflected at the system-reservoir interface. Also note, that we use a so-called partition-free method~\cite{ridley2018formal,zhang2013first}, meaning that the leads had been coupled to the system in a very distant past. The nonequilibrium situation at time zero is introduced only by the switching of the external magnetic field.   

The comparison in figure~\ref{fig:4_ComparisonOpenClosed}(b) confirms that the dynamics in a long isolated system can be to a large extent reproduced by a much shorter magnetic junction. 
We also show this in figure~\ref{fig:3_DampingJsd_Closed_2}(a), where the black circles represent effective Gilbert damping obtained from fitting the dynamics of a central system with only $N=11$ points coupled to leads. The results are in very good agreement with the data for long chains. In addition we show in Section~\ref{App:CvQ}, where we compare the dynamics of quantum and classical spins, that this technique can be used also to address dynamics of a large quantum spin coupled to a long chain.

\subsection{Influence of Nonequilibrium Electron Transport on Spin Dynamics}\label{sec:DCDependence}

The introduction of reservoirs allows us, besides the possibility to mitigate finite-size effects, to investigate the influence of nonequilibrium electron transport, driven by a finite voltage bias between the leads, on the classical spin dynamics.

We first consider a relatively long central system analogous to the above case, where we set $N=21$ and $\Gamma_{\ell}=2$. Afterwards, to clarify the role of particular system states on the spin relaxation, we address short systems $N=1$ and $N=5$ with a much smaller broadening $\Gamma_{\ell}=0.1$.   
In all scenarios, we initialize $\vec{S}_0 = \vec{e}_x$ and employ the partition-free method~\cite{ridley2018formal,zhang2013first} where the finite voltage drop was introduced in a distant past.
In practice, this means that we evolve the system with a fixed initial classical spin orientation until the steady state is reached. Only then, at $t=t_0$, the external field $\vec{B}=B\vec{e}_z$, perpendicular to the initial orientation of the spin, is switched on.

\begin{figure}[!h]
	\centering
	\includegraphics[width=1.0\columnwidth]{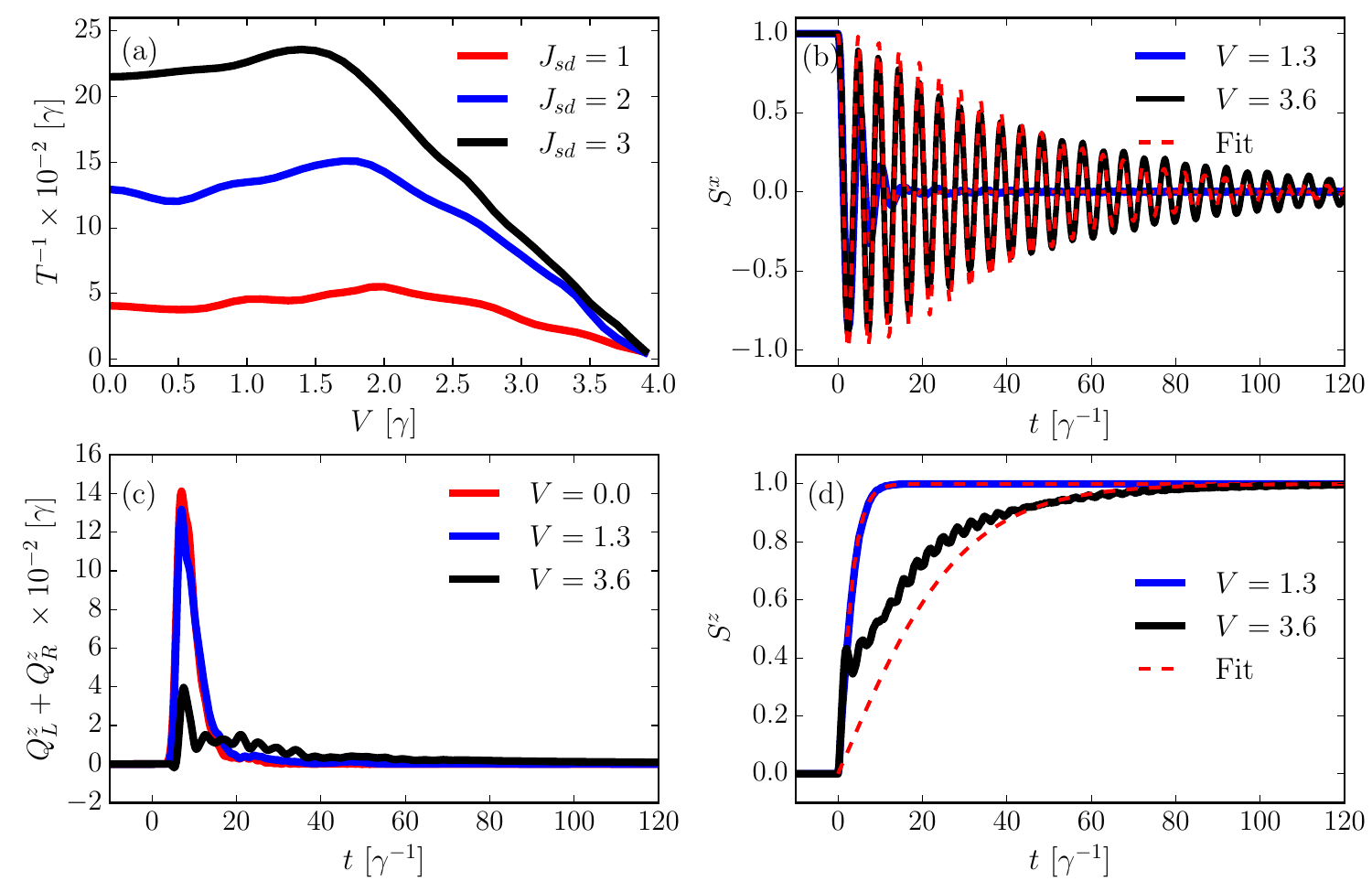}
	\caption{(a) Extracted relaxation rate $T^{-1}=\alpha b /(1+\alpha^2)$ as a function of voltage drop for $N=21$, $\Gamma_{\ell}=2$, $\beta_{\ell}=40$ and three values of $J_{sd}$. (b),(d) Examples of $S_{x}$ and $S_{z}$ dynamics for small and large voltages together with their LLG fits (red dashed lines). (c) Sums of spin currents which equal the spin torque experienced by the central spin.}\label{fig:chain}
\end{figure}

As in the above analysis of the closed system, we use equations~(\ref{eq:FitX})-(\ref{eq:FitZ}) to extract the effective damping and effective field, respectively, the relaxation rate. However, here this analysis has some important limitations. A typical comparison of the classical spin evolution and the corresponding LLG dynamics with fitted $\alpha$ and $b$ is illustrated in figure~\ref{fig:chain}(b,d), where we set $J_{sd}=3$, $N=21$ and $\Gamma_\ell=2$. In general, the fit of $S_x$ (or $S_y$) spin component dynamics is stable up to $V\approx 4$. Below this value, the voltage probes the energy window of the broadened states of the central system. Yet, the comparison of the time evolution of $S_z$ from the model equation~(\ref{eq:FitZ}), with $\alpha$ and $b$ fitted from the $x$-component of full evolution, with the full $S_z$ dynamics reveals that the simple LLG model is not correctly capturing the short time dynamics for large enough $V$. For $J_{sd}=3$ a clear deviation can already be seen at $V\approx 3$. This is related to the fact, that although the frequency of the dominant precession is stable, the real damping is actually time dependent. In addition,  the LLG with a constant effective $\alpha$ correctly describes the long-time spin dynamics only if we avoid the regime of long-living nutations, i.e., $J_{sd}>5$.  

Nevertheless, the fitting procedure can be used to capture the general qualitative behavior of relaxation represented by the effective relaxation rate $T^{-1}$ for small $J_{sd}$ and $V$. Figure~\ref{fig:chain}(a) shows dependencies of fitted rates $T^{-1}$ on the voltage for three values of $J_{sd}$. The relaxation rate is relatively stable for small voltages ($V<2$) with only small deviations from its equilibrium value. It rapidly declines above $V\approx 2$ and cannot be reliably extracted for $V\gtrsim4$. This can be partially explained by realizing that damping is determined by the dissipative processes of electrons leaving or entering the central system. 
These processes are proportional to the systems density of states at the Fermi energy of the reservoirs~\cite{nunez2008effective}. Therefore, the damping decreases with vanishing DOS. A crucial role for the damping plays the spin polarization of the relevant states and also the precession of the spin. These result in spin-polarized currents $\vec{Q}_\ell$ and spin-torques $\vec{Q}_{L} + \vec{Q}_\mathrm{R}$ acting on the classical spin. Although the overall charge current increases monotonically with $V$ (not shown here), the spin torque maxima diminish as illustrated in figure~\ref{fig:chain}(c).  

The slightly non-monotonic dependence of the relaxation rate on the voltage in figure~\ref{fig:chain}(a) reflects the details of the DOS of the central system. These effects of particular states are to a large extent overshadowed by the natural broadening from the leads. To elevate these details and to study them in a more controlled way, in the next section, we significantly lower the coupling to the leads and also reduce the central chain to one and then five sites. In this regime, the setup can be understood, e.g., as a simple model of a molecular magnet weakly coupled to metallic leads.

\subsection{Magnetic junction \label{Sec:MJ}}
\begin{figure*}[t]
	\centering
	\includegraphics[width=0.99\textwidth]{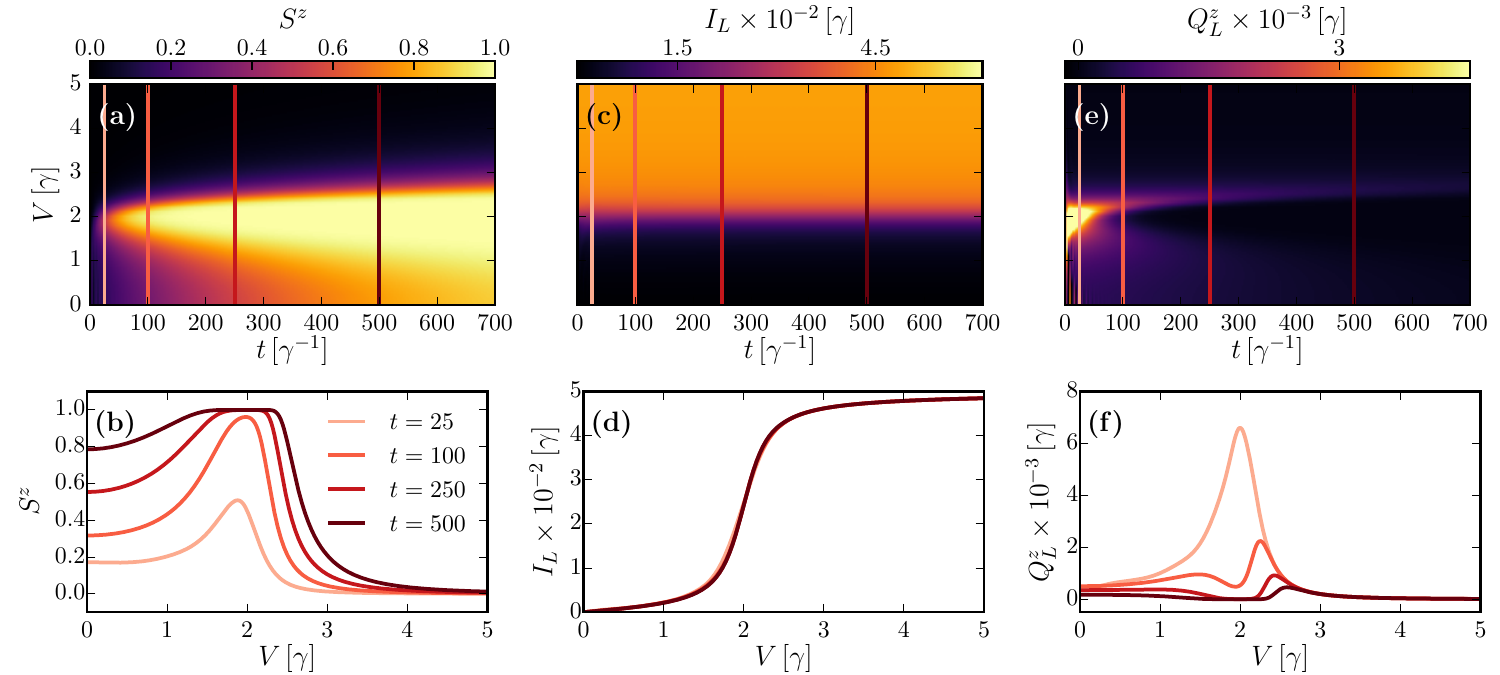}
	\caption{Time-evolution of $S^z(t,V)$ (a) and $S^z(V)$ at times $t\in\{25,100,250,500\}\gamma^{-1}$ (b), charge current $I_L(t,V)$ (c) and $I_L(V)$ (d), and spin current $Q^z_L(t,V)$ (e) and $Q^z_L(V)$ (f) measured at the left system-reservoir interface. The vertical lines in the first-row panels signal and color-code the time positions of $V$ dependencies in the second row. All panels show data for model parameters $\Gamma_\ell =0.1$, $J_{sd}=2.0$, $\beta_\ell = 40$ and $B=0.1$.}\label{fig:OpenSingleOverview}
\end{figure*} 

Figure~\ref{fig:OpenSingleOverview} shows the time evolution of the classical spin $S^z(t)$ (a),
the charge current $I_L(t)$ (c) and the spin current $Q^z_L(t)$ (e) measured at the left system-reservoir interface for various DC voltages $V$ in a single-site magnetic junction. 
Time slices (marked by vertical cuts of respective colors) taken for times $t\in\{25,100,250,500\}\gamma^{-1}$ are shown for $S^z$, $I_L$ and $Q^z_L$ in figure~\ref{fig:OpenSingleOverview} (b), (d), and (f), respectively.

To emphasize the role of particular states, we consider weak symmetric broadening functions $\Gamma_\ell \equiv \Gamma_L=\Gamma_R = 0.1$  and spin-electron coupling $J_{sd}=2$. This coupling is sufficient to significantly split the energy eigenvalues ($\varepsilon_\pm=\pm J_{sd}/2$) and simultaneously small enough to suppress the conditions which can therefore be neglected in the analysis of the spin dynamics.  
The external magnetic field was set to $B=0.1$. 

The results in figure~\ref{fig:OpenSingleOverview} (a),(b) reveal that spin relaxation is enhanced when the chemical potential of the reservoir $\ell$ matches one of the energy levels $\varepsilon_\pm$ of the central system. As before, this can be partially attributed to the elevated local density of states at these energies. The enhancement is caused by resonant tunneling of electrons between the reservoirs and the central energy level. To be more specific, the spin relaxation in this voltage regime is a two-step process: First, precession of the classical spin in the $x-y$ plane induces electronic spin excitations. 
Second, these excitations are transmitted into the reservoirs via polarized electron hopping, as seen, for example, from the enhanced transient spin current $Q^z_L$ at $V=2$, shown in figure~\ref{fig:OpenSingleOverview} (e),(f). Equivalent relaxation mechanisms have already been described in a number of studies addressing different structures, e.g., thin metal films, and are known as spin-pumping \cite{tserkovnyak2002enhanced,vsimanek2003gilbert,mills2003ferromagnetic,zutic2004spintronics,liu2014interface}.

On the other hand, spin relaxation is strongly suppressed for $\vert \mu_\ell\vert \ll \vert\varepsilon_\pm\vert$ and even more so for $\vert\mu_\ell\vert \gg \vert\varepsilon_\pm\vert$.
Because of the weak coupling to the leads, there is only a small charge current for small voltages, i.e., in the $\vert \mu_\ell\vert \ll \vert\varepsilon_\pm\vert$ regime. Consequently, the dynamic excitations generated by the classical spin in the electronic part are mostly localized in the central system as the tunneling into the reservoirs is limited.
On the contrary, for $\vert\mu_\ell\vert \gg \vert\varepsilon_\pm\vert$ both levels contribute to electronic transport [figure~\ref{fig:OpenSingleOverview}(c)] and we see large charge transport. However, both levels are populated almost equally $P_+(t)\approx P_-(t)$ which leads to a vanishing spin polarization $s^z(t) \sim P_+(t)-P_-(t) \approx 0$ and consequently to a strong suppression of spin relaxation. This is clear from the vanishing transient spin current $Q^z_L$ in this voltage-regime as shown in figure~\ref{fig:OpenSingleOverview}(e),(f).

Nevertheless, the aforementioned qualitative analysis neglects crucial mechanisms. It relies on the assumption that the system consists only of static energy levels $\varepsilon_\pm$, which holds true only for small magnetic fields $B$. However, for stronger magnetic fields,  one cannot neglect the fact that spin dynamics influences the electronic states~\cite{filipovic2013spin}. Specifically, $H_\mathrm{C}(t)$ varies with time due to the precession of the classical spin at a frequency $\omega_p$, which differs from the Larmor frequency as discussed in 
Refs.~\cite{stahl2017anomalous,bajpai2020spintronics,zhang2004roles,zhang2009generalization} and also in~\ref{App:Frequency}. 
The main resulting effect can be understood by assuming a (nearly) time-periodic classical spin $\vec{S}(t+2\pi/\omega_p) \approx \vec{S}(t)$ (which is fulfilled in the high-voltage regime). This introduces an essential effect whereby the oscillations generate frequency-dependent side bands, denoted by  $\overline{\varepsilon}_\pm = \varepsilon_\pm \mp |\omega_p|$ and $\tilde{\varepsilon}_\pm = \varepsilon_\pm \pm |\omega_p|$,  to the central levels $\varepsilon_\pm$~\cite{filipovic2013spin}. 
Two of these additional channels (in our case $\tilde{\varepsilon}_\pm$)  are available for transport via inelastic tunnel processes. For a more detailed analysis of such processes, we refer the interested reader to the work of Filipovi\'c \textit{et al.} in Ref.~\cite{filipovic2013spin}. It is important to emphasize that in the intermediate voltage regime, where the aforementioned assumption does not hold, a feedback mechanism between the conduction electrons and the classical spin leads to damping of spin dynamics. In this regime, the additional electronic states $\tilde{\varepsilon}_\pm$ should be understood as time-dependent. Technically, their time-dependence arises due to the non-perturbative character of our approach. 

\begin{figure}[!h]
	\centering
	\includegraphics[width=0.65\textwidth]{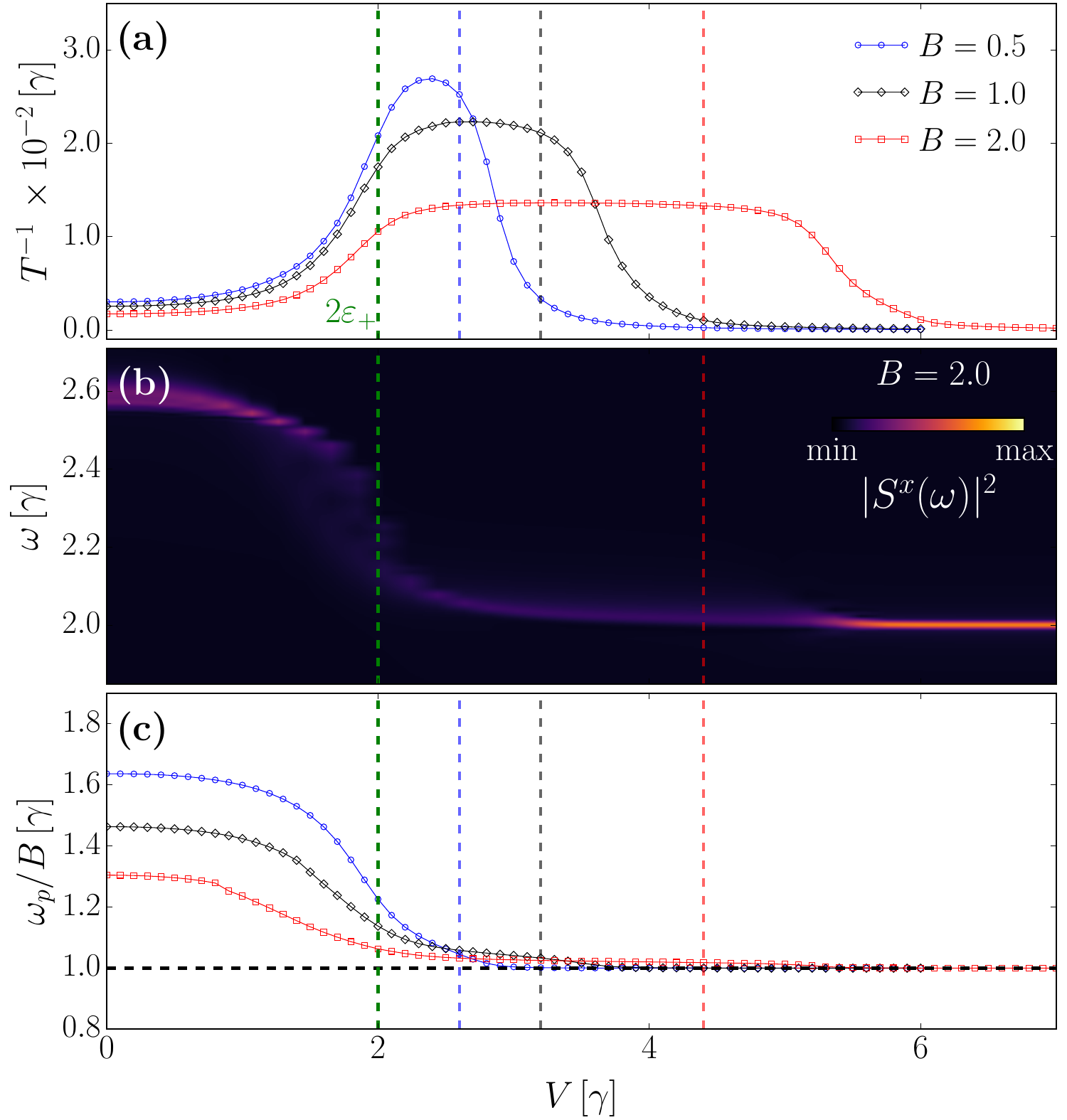}
	\caption{Relaxation rate $T^{-1}$ (a), spectral density $\vert S^x(\omega)\vert^2$ for $B=2$ (b) and dominant precession frequency $\omega_p$ (c) for various dc voltages $V$ and several field strengths $B$ in a hybrid dot with $J_{sd} = 3$, $\Gamma_\ell=0.1$, $\beta_\ell=40$. Dashed vertical lines denote $2\varepsilon_+$ (green), $2(\varepsilon_+ + \omega_p^{B}(V=0))$ for $B=0.5$ (blue), $B=1$ (black) and $B=2$ (red) and $\omega_p^B = 1.2 B$.}\label{fig:OpenSingleFreequency}
\end{figure}
Although these additional channels are of a transient nature, they have a significant influence on the voltage-dependence of spin-relaxation rates. They result in a magnetic field-dependent broadening of the voltage range in which the spin relaxation rates are enhanced. This effect is illustrated in figure~\ref{fig:OpenSingleFreequency}, where panel (a) depicts the influence of the external magnetic field strength ($B=0.5, 1, 2$) on the voltage-dependent relaxation rate $T^{-1}$. Consistent with previous settings, we fix $J_{sd} = 2$ and $\Gamma_\ell=0.1$. To quantify the precession frequency at different voltages, we analyze the spectral density $\vert S^x(\omega)\vert^2$ obtained through the expression
	\begin{equation}
		S^x(\omega) = \lim_{\tau \rightarrow \infty}\int_0^\tau dt \ S^x(t)e^{-i\omega t},
	\end{equation}
where in practice we, however, use a finite upper limit of the integration time $\tau$ set to $100\gamma^{-1}$, since on this timescale the classical spin is typically fully relaxed. This is exemplified in figure~\ref{fig:OpenSingleFreequency}(b) for $B=2$. Further details can be found in the discussion in~\ref{App:Frequency}. Moreover, we determine the dominant precession frequency $\omega_p$ by fitting it using equation~(\ref{eq:FitX}) as shown in figure~\ref{fig:OpenSingleFreequency}(c). The extracted frequency aligns well with the frequency obtained at the maximum of $\vert S^x(\omega)\vert^2$.
 
The vertical green dashed line in figure~\ref{fig:OpenSingleFreequency} signals the voltage at which the chemical potential of the left lead is aligned with $\varepsilon_+$. The other three vertical lines indicate positions of 
$2\tilde{\varepsilon}_+$. The results shown in figure~\ref{fig:OpenSingleFreequency} (a) affirm an enlargement of the voltage window in which spin relaxation is enhanced due to the side bands. 
As shown by the vertical dashed lines in figure~\ref{fig:OpenSingleFreequency}, the precession frequency $\omega_p^B=1.2B$ induced by the external field (for $J_{sd}=3$, obtained from the analysis shown in~\ref{App:BVar}) provides a quantitative bound on this voltage-window as $2\varepsilon_+ \leq V\leq 2(\varepsilon_+ +\omega_p^B)$ (shown in figure~\ref{fig:OpenSingleFreequency} by dashed lines, colored according to the respective $B$).
Figure~\ref{fig:OpenSingleFreequency} (c) shows that as the voltage is increased and reaches the regime $\vert \mu_\ell\vert \approx \vert \varepsilon_\pm\vert$, the dominant precession frequency gets strongly shifted towards the Larmor precession frequency $\omega_L$ but stays well above it until $\vert \mu_\ell\vert \lesssim 2\tilde{\varepsilon}_+$. Finally, in the high-voltage regime $\vert \mu_\ell\vert \gg \vert \varepsilon_\pm\vert$ the precession frequency is identical to $\omega_L$. 
This result further accentuates the previously mentioned decoupling of electrons from the classical spin in the large voltage regime.

\begin{figure}[!h]
	\centering
	\includegraphics[width=0.9\columnwidth]{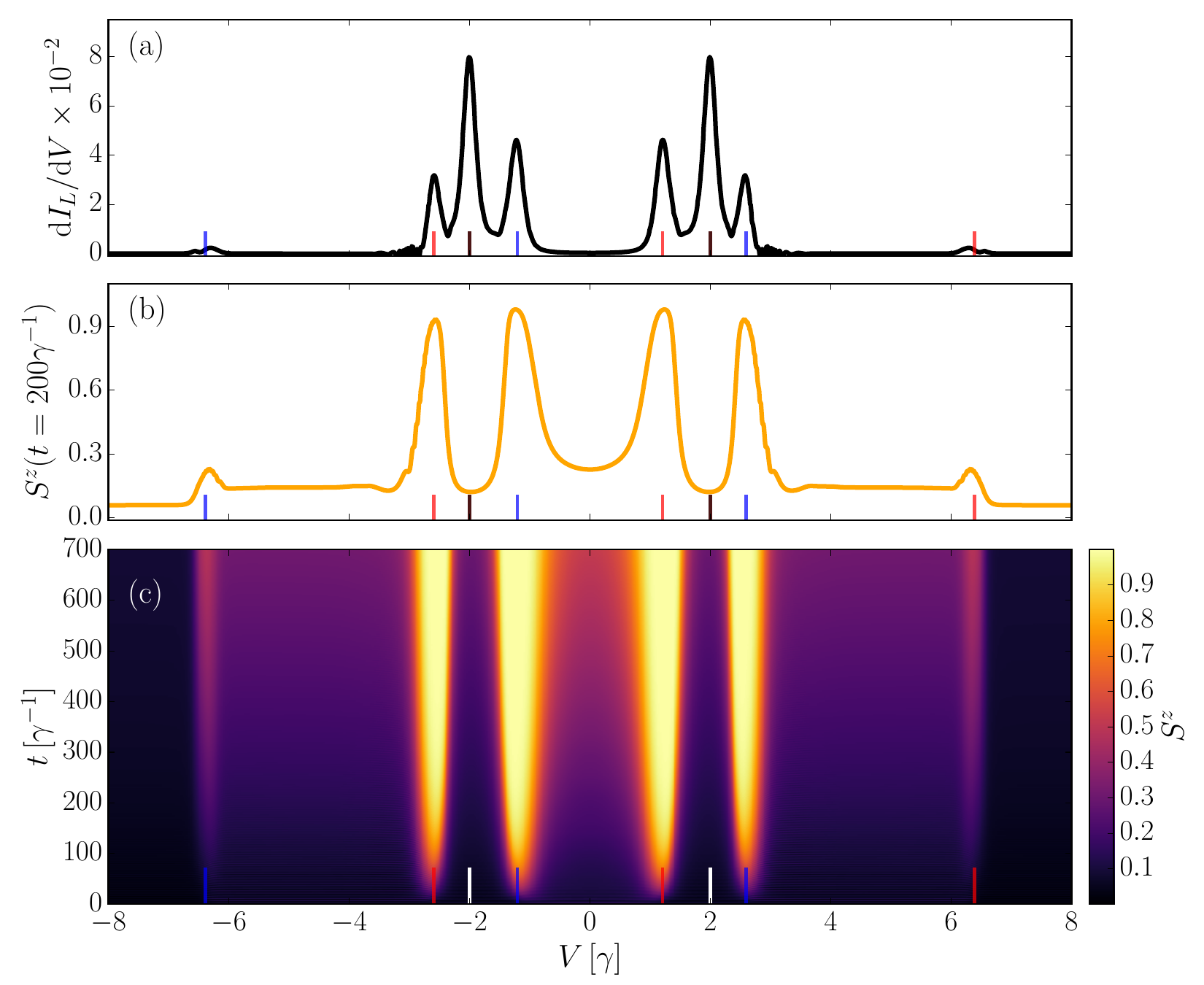}
	\caption{Charge-current conductance $\mathrm{d}I/\mathrm{d}V$ calculated at time $t=200\gamma^{-1}$ (a), snapshot of $S^z$ at time $t=200\gamma^{-1}$ (b) and the time-evolution of $S^z$ (c), all three as functions of voltage $V$. Parameters are $N=5$, $\Gamma_\ell = 0.1$, $J_{sd}  = 5$, $\beta_\ell=40$, and $B=0.1$.
	Eigenenergies of an isolated system are depicted by vertical bars, colored according to their averaged spin polarization $\langle s_n^z\rangle = \sum_k \langle \phi_n(\vec{r}_k)\vert \sigma^z\vert \phi_n(\vec{r}_k)\rangle$ of eigenstate $\vert \phi_n(\vec{r}_k)\rangle$ over all sites $k$: Red for $\langle s^z_n\rangle = +\frac{1}{2}$, blue for $\langle s^z_n\rangle = -\frac{1}{2}$, and degenerate states are depicted in black  in panels (a),(b) and white in (c).}\label{fig:MultiSite}
\end{figure}
Similar observations hold for central systems consisting of multiple sites with the classical spin at the center of the chain. Figure~\ref{fig:MultiSite} shows the differential conductance $\mathrm{d}I_L/\mathrm{d}V$ (a) and classical spin $z$-projection $S^z$ (b) both at time $t=200\gamma^{-1}$ and the time evolution of  $S^z$ (c), all three as functions of voltage drop $V$ for $N=5$, $\Gamma_\ell = 0.1$, $J_{sd} = 5$ and $B=0.1$. We use these directly obtained quantities to illustrate the resonance features instead of fitted rates $T^{-1}$ as the fitting with model equations~(\ref{eq:FitX}),(\ref{eq:FitZ}) proved to be unstable. 
 
Panels (b) and (c) in figure~\ref{fig:MultiSite} reveal that the relaxation is maximal when the chemical potential of one of the reservoirs is in resonance with some of the $2N$ energy levels $\varepsilon_n$ of $H_\mathrm{C}$ [plotted as color bars in the bottom part of panels figure~\ref{fig:MultiSite}(a)-(c)]. The analysis of a multi-site central system also shows that not all states contribute equally to spin relaxation. In particular, despite high conductance, the degenerate states ($\varepsilon_n=\varepsilon_m$ for $n\neq m$, black bars) contribute only weakly to spin relaxation. This can be deduced from the local minima in $S_z$ for $V=\pm 2$ although $\dif I/\dif V$ is maximal there. The degeneracy of these states leads to a vanishing local spin polarization. Therefore, these states do not couple to the classical spin and hence leave spin dynamics unaffected. 

Moreover, we observe that longer chains exhibit significantly enhanced suppression of relaxation between resonant features when compared to single-site systems. This suppression becomes apparent at energy values close to zero ($\vert V\vert \approx 0$) or at energy values approximately equal to $\pm 1$ ($\vert V\vert \approx 2$). The underlying cause can be attributed to a substantial decrease in the local density of states at corresponding energies, which stems from the rapid decline in broadening as the distance from the leads increases, particularly for strong $J_{sd}$ coupling~\cite{freericks2016transport, Zonda2019}. Additionally, as discussed in Section~\ref{App:LocEigenstates}, the localization of electrons also contributes to these processes.

\section{Summary}\label{sec:Summary}
We have investigated the relaxation dynamics of a hybrid quantum-classical spin system using the QC-EOM method. 
In a first step, we have analyzed spin dynamics in an isolated hybrid system, where we have extracted the relaxation rates and Gilbert damping by fitting numerical solutions to the classical LLG dynamics. The results for the damping as well as relaxation rates of the classical spin show a pronounced maximum at finite electron-spin coupling, which can be qualitatively understood by investigating the local density of states on the Fermi level. 
We compare our results with the prior work conducted by Sayad and Potthoff~\cite{sayad2015spin}.
We show that our results follow the quadratic scaling of the relaxation rate with $J_{sd}$ as predicted by approximate analytical solutions.  
This confirms the stability of our method even in the regime of small relaxation times, which proved to be difficult in the previous work. We also argue that the discrepancy in the position of its maxima is due to the differences in the extraction of the relaxation time between the two works.
We have also shown that the dynamics of a closed system with a very long chain can be reproduced by a short system coupled to infinite leads with a flat band.  

Extending the study to a magnetic nanojunction coupled to fermionic reservoirs and subjected to an external dc voltage reveals a strong influence of the bias voltage on spin relaxation. 
We have observed that a clear signature of the electronic spectrum is imprinted in the voltage-dependent spin dynamics in an non-trivial way where the spin relaxation is boosted only by the magnetically polarized single-particle eigenstates. The dynamical side bands induced to the electronic spectrum due to the precession boost the damping in a wide range of the voltage drop.  

We note that despite the fact that it is in good quantitative agreement with the full quantum approach if the spin is large, as briefly discussed in Sec~\ref{App:CvQ}, the used quantum-classical mode has its limitations~\cite{mondal2021can}. The classical representation of the localized spin cannot account for some inherently quantum phenomena. These include the Kondo effect~\cite{hewson1997kondo,sayad2015spin}, 
damping of nutations~\cite{sayad2016inertia} or possible torques caused by the many-body character of quantum states~\cite{mondal2019quantum,petrovic2021prx}. 
However, the formulation of the QC-EOM as a hierarchical master equation invites its future generalizations to a fully quantum system by going beyond the second tier in the expansion~\cite{zheng2009numerical,thoss2018perspective}. 

\section{Acknowledgements}
The authors acknowledge support by the state of Baden-Württemberg through bwHPC
and the German Research Foundation (DFG) through grant no INST 40/467-1 FUGG (JUSTUS cluster). 
R.S. and M.T. acknowledge support by DFG-funded Research Training Group DynCAM (Grant No. RTG 2717).
M.\v{Z}. acknowledges support by the Czech Science Foundation via Project No. 22-22419S. 

\appendix
\section{Current matrices \label{App:CurMat}}

As shown, e.g., in Ref.~\cite{haug2008quantum} the current matrices can be calculated as   
\begin{equation}
	\Pi_\ell (t) = \int_{-\infty}^{\infty} \dif \tau [G^<(t,\tau)\Sigma_\ell^a(\tau,t) G^r(t,\tau)\Sigma_\ell^>(\tau,t)],
\end{equation}
where $G^<(t,\tau)$ is the lesser nonequilibrium Green function, $G^r(t,\tau)$ is the retarded nonequilibrium Green function and $\Sigma_\ell^{a/<}(\tau,t)$ is the advanced/lesser self-energy due to the coupling to the leads. Following Croy and Saalmann~\cite{croy2009propagation} this expression can be rewritten using the general relation valid for both Green functions and self-energies
\begin{equation}
 X^{a,r}(t,t')=\pm\Theta(\pm t\mp t')\left[X^>(t,t')-X^<(t,t'))\right],
 \label{eq:XAR}
\end{equation} 
to the current matrix form as stated in equation~(\ref{eq:CM}) where $G^<_{\alpha\beta}(t,\tau) = i\langle c_\beta^\dagger(\tau)c_\alpha^\pdagger(t)\rangle$ and $G^>_{\alpha\beta}(t,\tau) = -i\langle c_\beta^\pdagger(t)c_\alpha^\dagger(\tau)\rangle$. The related equation of motion, given in  equation~(\ref{eq:PadeAuxiliaryRhoEOM}), can be derived using the Kadanoff-Baym relations
\begin{eqnarray}
	i&&\frac{\partial}{\partial t}G^\lessgtr(t,t')=H_C(t)G^\lessgtr(t,t')+\\
	&&\int d\tau \left[\Sigma^r(t,\tau)G^\lessgtr(\tau,t')+\Sigma^\lessgtr(t,\tau)G^a(\tau,t')\right],
\end{eqnarray}    
by using again the expressions in equation~(\ref{eq:XAR}) together with the identity $\rho_{\alpha\beta}(t)=-iG^<_{\alpha\beta}(t,t)$. In our case we further simplify the expression by assuming the wide-band limit and time independent coupling of the system to the leads as stated in equation~(\ref{eq:Gamma}), meaning that the leads have been always coupled to the central system (partition-free method). In addition, the finite voltage drop is introduced in distant past $t_v \ll t_0$ and we initially evolve the system with fixed classical spin till the steady state is reached. In principle any other protocol including time-dependent voltage where $\chi_{p\ell}^+ (t) = \mu_\ell(t)+ i\xi_p \beta^{-1}$, can be treated by the method. A detailed derivation of such and similar cases (with different Pad\'{e} coefficients as used in the present work) can be found in Ref.~\cite{croy2009propagation}.

\section{Influence of $s$-$d$ Coupling on Spin Precession}\label{App:Frequency}
\begin{figure}[!h]
	\centering
	\includegraphics[width=0.65\columnwidth]{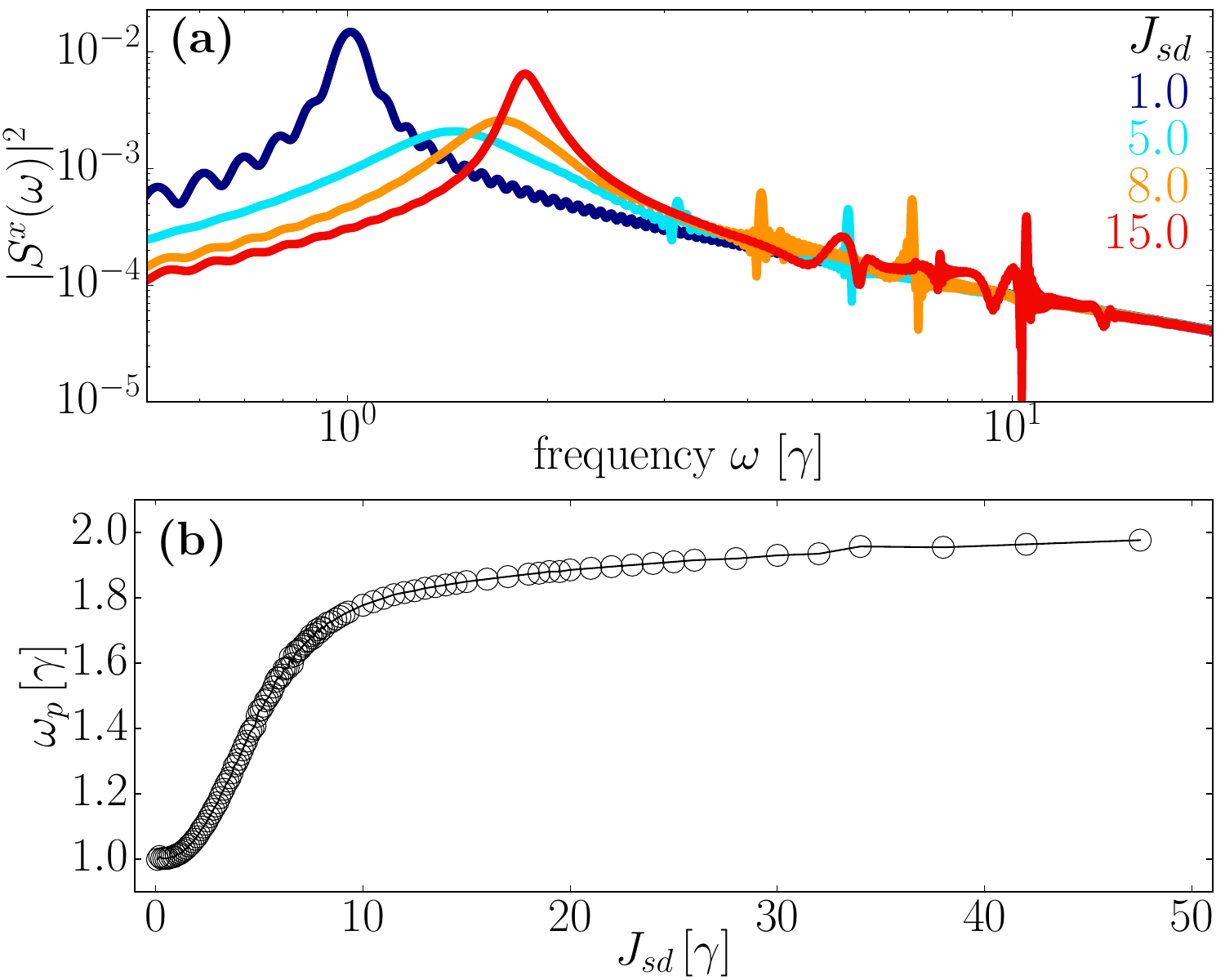}
	\caption[Spectral density and spin-precession frequency for various $J_{sd}$.]{Spectral density of the classical spin $\vert S^x(\omega)\vert^2$ obtained from the Fourier-transform of $S^x(t)$ for selected $J_{sd}$ (a) and therefrom obtained dominant precession frequency $\omega_p$ for various $J_{sd}$ (b).}\label{fig:3_SpinPol_SpectralDens}
\end{figure}

Here we address some additional details of the influence of $J_{sd}$ on the spin-dynamics, in particular, on the spin precession frequency $\omega_p$. In accordance with the case presented in the main text in Sec.~\ref{sec:Closed} we set
the chain length to $N=151$ and $B=1$. We show the spectral density  $\vert S^x(\omega)\vert^2$ (a) for various $J_{sd}$ and the dominant frequency as a function of $J_\textrm{sd}$ (b) in figure~\ref{fig:3_SpinPol_SpectralDens}.  
Finite $J_{sd}$ shifts the dominant precession frequency away from the Larmor value $\omega_p=B$ [panel (b)] which was already discussed by Stahl and Potthoff~\cite{stahl2017anomalous} and others, e.g.,  ~\cite{bajpai2020spintronics,zhang2004roles,zhang2009generalization}. 
From our analysis, we obtain the shifted precession frequency due to spin-electron coupling as $\omega_p^{B} = 1.2B$ for $J_{sd}=3$ as employed in Sec.~\ref{sec:DCDependence}. 

In addition, strong $J_{sd}$, e.g. $J_{sd}=15$ in panel (a), also induces distinct high-frequency oscillations. 
The high-frequency oscillations can be attributed to higher-order terms in spin dynamics, e.g., by assuming a Taylor-series expansion
\begin{equation}
	\dot{\vec{S}} \sim \vec{S}\times \sum_\alpha c_\alpha \frac{\dif^\alpha \vec{S}}{\dif t^\alpha}.
\end{equation}
Notable terms are spin-precession $\sim \vec{S}\times \vec{C}$ with constant vector $\vec{C}$ (peak-position $\omega_p$ of $\vert S^x(\omega)\vert^2$), Gilbert-damping $\sim \vec{S}\times \dot{\vec{S}}$ (broadening $\Gamma_{S^x(\omega)}$ of the peaks of $\vert S^x(\omega)\vert^2$), and inertia $\sim \vec{S}\times \ddot{\vec{S}}$ giving rise to nutation on a short time-scale \cite{hammar2017transient,thonig2015gilbert,stahl2017anomalous}. The observed fast oscillations, and, hence, the high-frequency peaks in figure~\ref{fig:3_SpinPol_SpectralDens} (a), can be attributed to nutation or even higher-order effects. Due to the increasing significance of these contributions with increasing $J_{sd}$, we assess electronic dynamics as the root cause for these fast oscillations~\cite{hammar2016time}.  

\section{Effect of External Field Strength on Classical Spin Relaxation Rate}\label{App:BVar}
\begin{figure}[!h]
	\centering
	\includegraphics[width=0.9\textwidth]{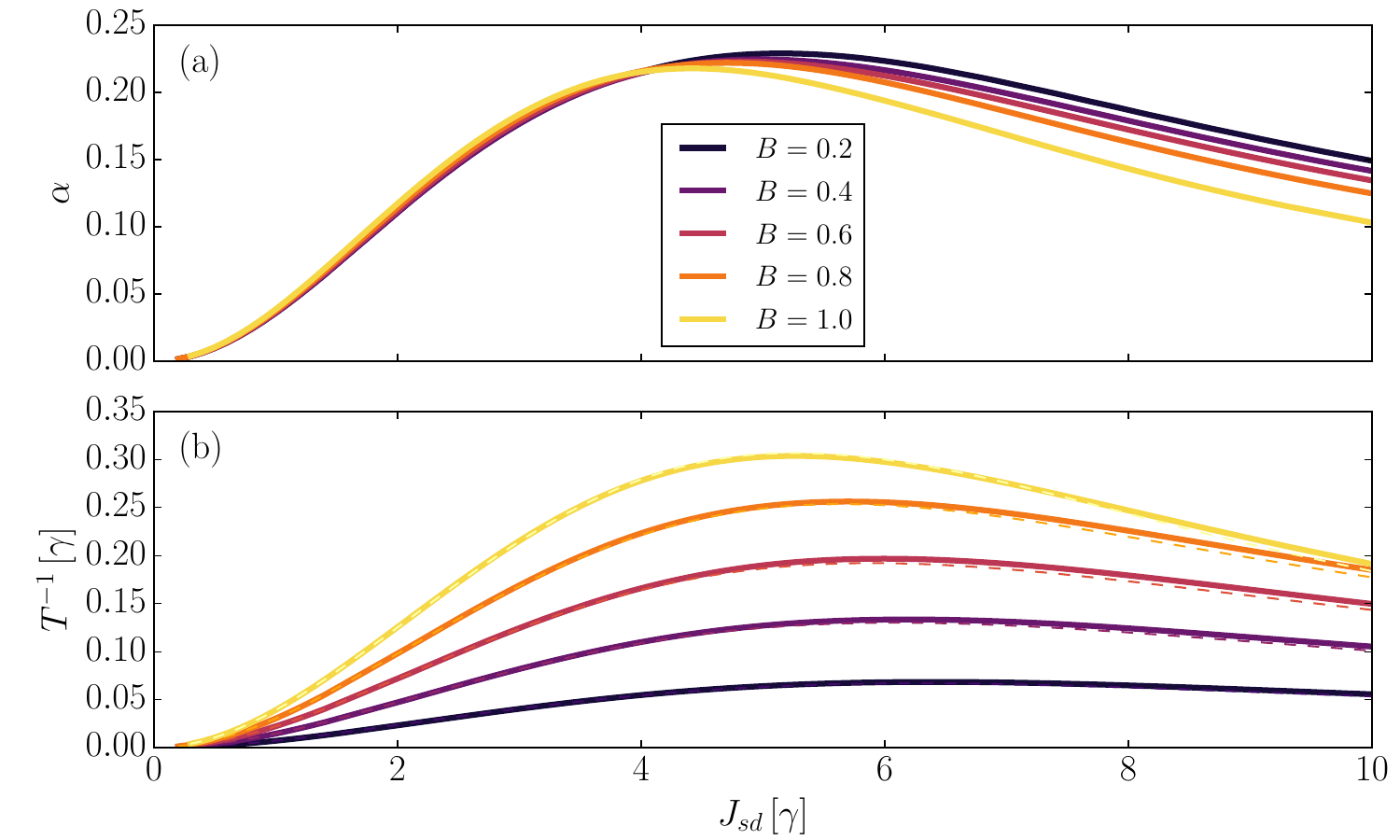}
	\caption{Effective Gilbert damping $\alpha$ (a) and relaxation rates $T^{-1}$ (b), both as functions of $J_{sd}$ and for different field strengths $B$ in a classical single-impurity Kondo chain with $N=151$. The full line in (b) shows the relaxation rate obtained from $S^z$, whereas the dashed line the one obtained from $S^x$. Variance due to the fits in (b) is smaller than the line widths.}\label{fig:Relax_BVar}
\end{figure}

In Sec.~\ref{sec:Closed} we identify the three main $J_{sd}$ coupling regimes, which shows that spin relaxation is significantly influenced by the $s$-$d$ coupling. These results were obtained for $B=1$, whereas here we show that these observations hold for various $B$ as well. 
Figure~\ref{fig:Relax_BVar} (a) shows the effective Gilbert $\alpha$ for various $J_{sd}$ and $B$ and figure~\ref{fig:Relax_BVar} (b) the therefrom obtained relaxation rate $T^{-1}$, using $S^z$ (full lines) and $S^x$ (dashed lines). For all external field strengths, the dependence of the relaxation rate on $J_{sd}$ shows a similar characteristic and they are mostly independent of whether fitting $S^x$ or $S^z$ as can be seen from the similar $J_{sd}$ dependence of the full and dashed lines for all investigated field strengths $B$. 

\section{Localization Properties of Single-Particle Eigenstates}\label{App:LocEigenstates}
\begin{figure}[!h]
	\centering
	\includegraphics[width=0.65\textwidth]{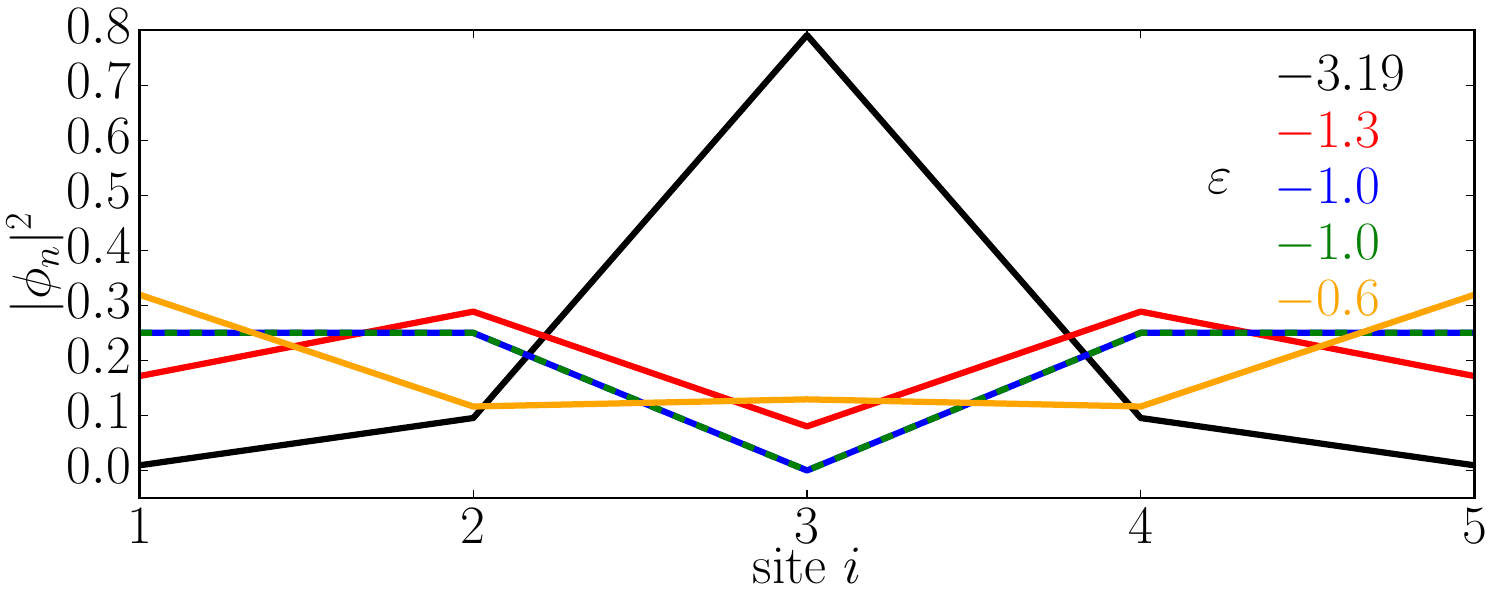}
	\caption{Space-resolved single-particle eigenstates $\vert\phi_n\rangle$ of Hamiltonian given in equation~\ref{eq:Hamiltonian_Quantum_Closed} with energy $\varepsilon_n$ with same parameters as in figure~\ref{fig:MultiSite}, $N=5$, $J_{sd}=5$ and $\mu=0$, for a classical spin $\vec{S}=\vec{e}_z$ positioned at the center of the chain.}\label{fig:App_SingleParticleEigenstates}
\end{figure}

In Sec.~\ref{Sec:MJ}, we argue that certain properties of single-particle eigenstates significantly influence voltage-dependent spin-dynamics in a magnetic junction. To support this statement, we show in figure~\ref{fig:App_SingleParticleEigenstates} the site-resolved amplitude $\vert\phi_n\vert^2$ of single-particle eigenstates $\vert\phi_n\rangle$ corresponding to eigenenergies in the lower band $\varepsilon_n<0$, in a closed chain with $N=5$ with $J_{sd}=5$.
The state with energy $\varepsilon = -3.19$ (figure~\ref{fig:App_SingleParticleEigenstates} black line) is clearly localized around site $j=3$ and its probability density rapidly decays with increasing distance from this site. All other states, on the other hand, show a less pronounced spatial distribution. Degenerate states (e.g., states corresponding to eigenenergy $\varepsilon=-1$, blue and green line), have vanishing probability density at the site of the classical spin.
Thus, these states have a vanishing contribution to nonequilibrium spin dynamics when tuning the reservoir chemical potentials in resonance with the corresponding eigenenergies of these states although the electronic transmission function for these energies is finite. 

\section{Derivation of Hybrid Spin Equation of Motion}\label{App:HybridSpinEOM}
\indent In this appendix we derive the effective spin equation of motion within the QC-EOM approach.
We start with the differentiation of $\vec{s}(t)$ (equation~\ref{eq:SpinPolarization}) with respect to time and employ the hierarchical equations of motion (equation~\ref{eq:RhoOpenEOM}), resulting in
\begin{eqnarray}
2\frac{\dif \vec{s}_m(t)}{\dif t} &=& \mathrm{tr}\left\{\frac{\dif \rho_m}{\dif t}\bm{\sigma}\right\}\label{eq:4_SpinPOlDerivative}\\
&=& \mathrm{tr}\left\{\mathrm{tr}_{\Lambda\diagdown m}\left\{\frac{\dif \rho}{\dif t}\right\}\bm{\sigma}\right\}\\
&=&\mathrm{tr}\left\{-i([H_\mathrm{C},\rho])_m\bm{\sigma}\right\}  \label{eq:4_SpinPOlDerivative_1}\\
&&\qquad - \sum_\ell\left( (\vec{Q}^*_{\ell})_m +(\vec{Q}_{\ell}^{\phantom{*}})_m\right),
\label{eq:4_SpinPOlDerivative_2}
\end{eqnarray}
with the central system Hamiltonian $H_\mathrm{C}$ given in equation~(\ref{eq:Hamiltonian_Quantum_Closed}), the spin-current $\vec{Q}_\ell$ from equation~(\ref{eq:DefSpinCurrent}), and noting that the Pauli-matrices are hermitian.
Here, we use the notation $(A)_m = \mathrm{tr}_{\Lambda\diagdown m}\{A\}$
for operators $A$ in the reduced $2\times 2$ space at site $m$.
The first term on the right-hand side of equations~(\ref{eq:4_SpinPOlDerivative_1})-(\ref{eq:4_SpinPOlDerivative_2}) can be understood as the unitary (isolated system) or adiabatic time-evolution of the spin-polarization since
\begin{eqnarray*}
	\mathrm{tr}\{-i([H_\mathrm{C},\rho])_m\bm{\sigma}\} &=& \mathrm{tr}\left\{\frac{\mathrm{d} \rho^\mathrm{iso}_m}{\mathrm{d} t}\bm{\sigma}\right\}\\
	& =& \frac{\mathrm{d}}{\mathrm{d} t}\mathrm{tr}\{\rho^\mathrm{iso}_m\bm{\sigma}\}\\
	& =& \frac{\mathrm{d}\vec{s}^\mathrm{iso}_m}{\mathrm{d} t}
\end{eqnarray*}
The formal solution of equation~(\ref{eq:4_SpinPOlDerivative}) is found then
\begin{eqnarray}
\vec{s}_m(t) = \vec{s}_m(t_0) &-& \frac{1}{2}\int_{t_0}^t\dif \tau \bigg( \mathrm{tr}\{i([H_\mathrm{C}(\tau),\rho(\tau)])_m\bm{\sigma}\}\\
&+&\sum_\ell[(\vec{Q}_\ell(\tau)^*)_m+(\vec{Q}_\ell(\tau))_m]\bigg) .
\end{eqnarray}
The resulting equation of motion for the localized spin is then obtained using the equation of motion 
\begin{equation}
	\frac{\mathrm{d}}{\mathrm{d} t}\vec{S}(t) = J_{sd}\vec{s}_m(t)\times \vec{S} - \vec{B}\times \vec{S},
\end{equation}

with $\vec{s}_m(t) = \mathrm{tr}_{\Lambda\diagdown \{m\}}\{\vec{s}(t)\}$. Therefrom follows the exact equation of motion
\begin{eqnarray}
\frac{\dif \vec{S}}{\dif t} 
&=&\bigg(J_{sd}\vec{s}_m(t_0)-\vec{B}\bigg)\times\vec{S}(t)\\
&&\qquad +J_{sd}\vec{S}(t)\times\int_{t_0}^t\dif \tau\,\mathrm{tr}\{i([H_\mathrm{C}(\tau),\rho(\tau)])_m\bm{\sigma}\}\\
&&\qquad +J_{sd}\sum_\ell \vec{S}(t)\times\int_{t_0}^t\dif \tau ((\vec{Q}_\ell^*(\tau))_m+(\vec{Q}_\ell(\tau))_m).
\label{eq:4_SpinEOM_Appendix}
\end{eqnarray}
Note, that due to the integral over the full history of the system, contained in the time-dependence of $H_\mathrm{C}(\tau)\equiv H_\mathrm{C}[\vec{S}(\tau);\tau]$, the resulting equation of motion for $\vec{S}$ takes that of a non-Markovian Master equation.

\section{Classical versus quantum spin}\label{App:CvQ}
\begin{figure}[!h]
	\centering
	\includegraphics[width=0.85\textwidth]{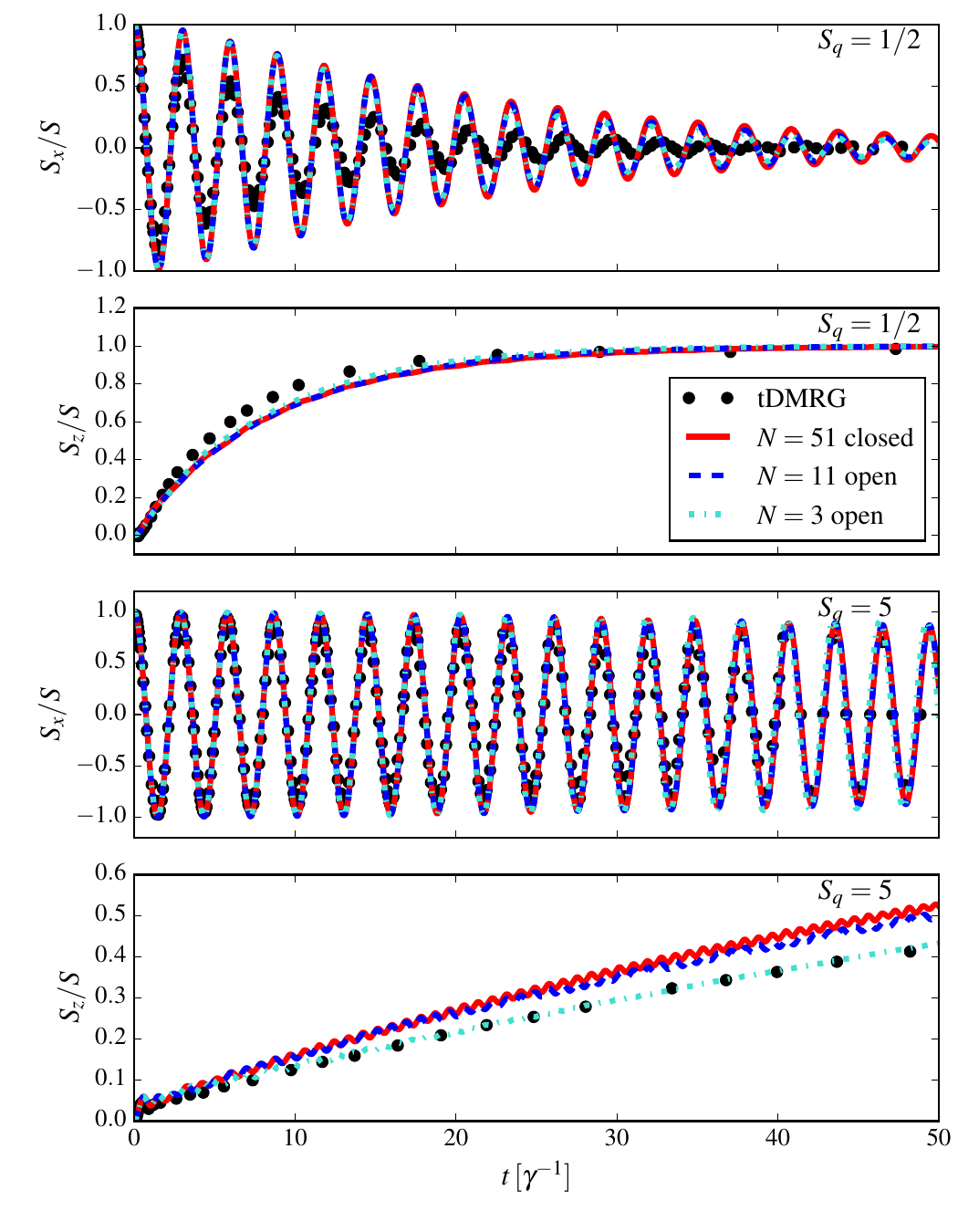}
\caption{Comparison of the transient spin dynamics of a single spin coupled sideways (left edge) to a fermionic chain with $J_{sd}=1$. The spin was initially prepared in the $x$ direction, and at time zero a magnetic field $B_z=1$ in the $z$ direction was switched on. (a), (b) Time evolution of the $S_x$ and $S_z$ components of the spin. Here, the dots represent tDMRG results for a quantum spin with $S_q=1/2$ coupled to a chain with 50 sites taken graphically from Ref.~\cite{sayad2016inertia}. The lines show our QC-EOM results for $S =\sqrt{S_q(S_q+1)}=\sqrt{3/4}$. Here the red solid line represents a closed system with 51 sites, the dashed blue lines open system with $N=11$ sites and the turquoise dashed line represents an open system with three sites. In open systems, the chains are coupled to a lead, treated in the WBL with $\Gamma_R=2$, at the opposite end of the chain  ‌to where the spin is located. Panels (c) a (d) show the same as (a) and (b) but for $S_q=5$ and $S=\sqrt{5(5+1)}=\sqrt{30}.$  }\label{fig:App_CvQ}
\end{figure}

Here we briefly address the difference between the dynamics of a classical spin and that of a quantum spin coupled to a fermionic chain. To this end, we focus on a system previously investigated by Sayad et al. in Ref.~\cite{sayad2016inertia}, namely a classical or quantum spin coupled sideways (at the left edge) to the chain of otherwise non-interacting electrons. This system can be described by the Hamiltonian~(\ref{eq:Hamiltonian_Quantum_Closed}) with $m=1$ and where $\vec{S}_1$ is either a quantum spin characterized by the quantum number $S_q$ or a classical spin with fixed length $|\vec{S}_1|=\sqrt{S_q(S_q+1)}\equiv S$. We investigated dynamics where the spin was initially prepared in the $x$ direction and the dynamics is triggered at $t=0$ by introducing a magnetic field $\vec{B}$ that points along the $z$ axes with $B_z=2$. In Fig.~\ref{fig:App_CvQ} we show two examples for $S_q=1/2$  ($S=\sqrt{3/4}$) in panels (a), (b); and for a large spin $S_q=5$ ($S=\sqrt{30}$) in (c), (d). Black circles represent the result for the localized quantum spin obtained by the time-dependent density matrix renormalization group technique (tDMRG) taken graphically from Fig.~1 in Ref.~\cite{sayad2016inertia}. The lines show our results for the classical spin in three different setups. The red lines represent the time evolution of the $x$ (a,c) and $z$ (b,d) components for the closed system where the classical spin is coupled to a tight binding chain of $N=51$ sites. The dashed blue line describes the time evolution of a classical spin coupled to chain of $N=11$ sites, which is on the right edge coupled to a WBL electronic reservoir with $\Gamma_R=2$. Finally, the turquoise line shows the system with $N=3$ coupled to the same type of lead as in the previous case.

Fig.~\ref{fig:App_CvQ} shows that the spin dynamics of the quantum spin with a large quantum number, represented here by $S_q=5$, is approachable by that of a classical spin of length $S = \sqrt{S_q(S_q+1)}$. Moreover, one can address this dynamics even with a short chain if coupled to an electronic reservoir in a way that mitigates the finite-size effects. On the other hand, as expected, the agreement between the QC and quantum approach decreases with the decreasing spin-quantum number. The $S_q=1/2$ case is in this respect the most extreme case, as it represents in a sense the "most quantum" case. The quantum case shows stronger damping at $S_q=1/2$ than the quantum-classical one. This is related to a more pronounced polarization of the spin density of the electrons at the site $1$, as discussed in more detail in Ref.~\cite{sayad2016inertia}.
\clearpage
\pagebreak
\section*{References}
\bibliographystyle{iopart-num}
\providecommand{\newblock}{}

\end{document}